\documentclass[useAMS,usenatbib]{mn2e}

\usepackage[dvips]{graphicx}
\usepackage{graphics}
\usepackage{lscape}
\usepackage{amssymb}
\usepackage{mathrsfs}
\input{epsf}

\newcommand\sersic{\,S\'{e}rsic}

\title[Morphological and Colour Effects of AGNs]
      {The Effects of an AGN on Host Galaxy Colour and Morphology Measurements}

\author[C.\ M.\ Pierce et al.]{
  \parbox[t]{\textwidth}{
  C.\ M.\ Pierce\thanks{E-mail: christina.pierce@physics.gatech.edu}$^{ \ 1,2}$,
  J.\ M.\ Lotz$^{3,4}$,
  J.\ R.\ Primack$^{5}$,
  D.\ J.\ V.\ Rosario$^{6}$,
  R.\ L.\ Griffith$^{7}$, \\
  C.\ J.\ Conselice$^{8}$,
  S.\ M.\ Faber$^{6}$,
  D.\ C.\ Koo$^{6}$,
  A.\ L.\ Coil$^{9}$,
  S.\ Salim$^{3}$, \\
  A.\ M.\ Koekemoer$^{10}$,
  E.\ S.\ Laird$^{11}$,
  R.\ J.\ Ivison$^{12}$ and
  R.\ Yan$^{13}$
  }
\vspace*{5pt} \\
$^{1}$Department of Physics, University of California, Santa Cruz, 1156 High Street, Santa Cruz, CA 95064, USA \\
$^{2}$School of Physics, Georgia Institute of Technology, 837 State Street, Atlanta, GA 30332-0430, USA \\
$^{3}$National Optical Astronomical Observatories, 950 N.\ Cherry Avenue, Tucson, AZ 85719, USA \\
$^{4}$Leo Goldberg Fellow \\
$^{5}$Santa Cruz Institute of Particle Physics, University of California, Santa Cruz, 1156 High Street, Santa Cruz, CA 95064, USA \\
$^{6}$UCO/Lick Observatory; Department of Astronomy and Astrophysics, University of California, Santa Cruz, 1156 High Street, \\ \hspace{0.9mm} Santa Cruz, CA 95064, USA \\
$^{7}$Jet Propulsion Laboratory, Caltech, MS 169-327, 4800 Oak Grove Drive, Pasadena, CA 91109, USA \\
$^{8}$School of Physics and Astronomy, University of Nottingham, Nottingham, NG7 2RD, UK \\
$^{9}$Department of Physics, University of California, San Diego, CA 92093, USA \\
$^{10}$Space Telescope Science Institute, 3700 San Martin Drive, Baltimore, MD 21218, USA \\
$^{11}$Astrophysics Group, Imperial College London, Blackett Laboratory, Prince Consort Rd., London SW7 2AW, UK \\
$^{12}$UK Astronomy Technology Centre, Edinburgh EH9 3HJ, UK \\
$^{13}$Department of Astronomy and Astrophysics, University of Toronto, 50 St.\ George Street, Toronto, ON M5S 3H4, Canada
}

\begin{document}

\date{Accepted 2010 February 09. Received 2010 February 04; in original form 2009 October 05}

\pagerange{\pageref{firstpage}--\pageref{lastpage}} \pubyear{2010}

\maketitle

\label{firstpage}

\begin{abstract}
We assess the effects of simulated active galactic nuclei (AGNs) on the colour and morphology measurements of their host galaxies. To test the morphology measurements, we select a sample of galaxies not known to host AGNs and add a series of point sources scaled to represent specified fractions of the observed $V$ band light detected from the resulting systems; we then compare morphology measurements of the simulated systems to measurements of the original galaxies. AGN contributions $^{>}_{\sim} 20$ per cent bias most of the morphology measurements tested, though the extent of the apparent bias depends on the morphological characteristics of the original galaxies. We test colour measurements by adding to non-AGN galaxy spectra a quasar spectrum scaled to contribute specified fractions of the rest-frame $B$ band light detected from the resulting systems. A quasar fraction of 5 per cent can move the NUV$-r$ colour of an elliptical galaxy from the UV-optical red sequence to the green valley, and 20 per cent can move it into the blue cloud. Combining the colour and morphology results, we find that a galaxy/AGN system with an AGN contribution $^{>}_{\sim} 20$ per cent may appear bluer and more bulge-dominated than the underlying galaxy. We conclude that (1) bulge-dominated, E/S0/Sa, and early-type morphology classifications are accurate for {\it red} AGN host galaxies and may be accurate for blue host galaxies, unless the AGN manifests itself as a well-defined point source; and (2) although highly unobscured AGNs, such as the quasar used for our experiments, can significantly bias the measured colours of AGN host galaxies, it is possible to identify such systems by examining optical images of the hosts for the presence of a point source and/or measuring the level of nuclear obscuration.
\end{abstract}


%%%%%%%%%%%%%%%%%%%%%%%%%%%%%
%%% Sections %%%
%%%%%%%%%%%%%%%%%%%%%%%%%%%%%
\section{Introduction}\label{intro}
A galaxy's colours and morphological characteristics supply clues about its star formation and interaction histories. For example, optical colours allow one to estimate the ages of observed stellar populations (e.g., Kennicutt 1998), while UV-optical colours are remarkably sensitive to recent star formation (e.g., Wyder et al.\ 2007). Morphological characteristics provide information about a galaxy's interaction history, and when combined with the recent star formation history they can be used to estimate the amount of time that has passed since the most recent galaxy merger (e.g., Lotz et al.\ 2008b). This is particularly useful for investigating potential causes of significant black hole growth (active galactic nuclei; AGNs), and possible connections between galaxy interactions and AGNs have been explored using both theoretical and observational techniques.

Many current simulations and semi-analytic models of galaxy interactions and evolution incorporate the effects of central supermassive black holes in the initial galaxies (e.g., Kauffmann \& Haehnelt 2000; Hopkins et al.\ 2005a,b, 2008a,b; Croton et al.\ 2006; Ciotti \& Ostriker 2007; Somerville et al.\ 2008). The results of these simulations invariably predict that the black holes become active during some stage of a merger event between gas-rich galaxies of similar masses (a `major merger', typically defined as a merger between galaxies having a mass ratio between 1:1 and 3:1). Most of these authors additionally found that some form of radiative output (or `feedback') from the AGN significantly decreased the star formation rate in the galaxy. Hopkins et al.\ (2005a,b, 2008a,b) described a form of AGN feedback that forcefully removes gas from the nuclear regions of a galaxy, after significant star formation and accretion onto the black hole has used up much of the gas. Croton et al.\ (2006) presented a complementary scenario, a `radio mode', in which AGN feedback heats gas in the nuclear regions to temperatures sufficient to stop star formation, without necessarily removing the gas from the galaxy.

Schawinski et al.\ (2007) and Georgakakis et al.\ (2008) described recent observational support for the scenarios derived from simulations (such as those summarized above) and semi-analytic models (e.g, Somerville et al.\ 2008). After visually examining $\sim$16,000 low-redshift early-type galaxies and spectroscopically classifying each as experiencing star formation, black hole growth, a combination of the two, or neither, Schawinski et al.\ (2007) presented evidence that radiative output from an AGN may cause its host galaxy to transition from a star-forming system to a quiescent system. They reported that the optical colours of the emission-line systems (20 per cent of their sample) are typically bluer than the quiescent systems, and the magnitude of the offset from the red sequence toward the blue cloud depends on the type of activity observed. Most of the systems dominated by emission from low-ionization nuclear emission line regions (LINERs) joined the inactive galaxies populating the red sequence, while the remaining LINER and Seyfert host galaxies typically exhibited colours on the red edge of the green valley. Early-type systems undergoing star formation were found in the blue cloud, and the systems exhibiting emission lines characteristic of both an AGN and star formation typically populated the green valley.

Georgakakis et al.\ (2008) stacked the X-ray emissions of three samples of galaxies at $z \sim 0.7$. The samples were comprised of galaxies with optical colours placing them on the red sequence, the green valley, or the blue cloud, and were limited to systems from which X-rays had not been detected at the limits of the survey. After stacking, they found significant X-ray detections from the red sequence and green valley samples, but not the blue cloud, indicating the presence of a significant population of obscured AGNs among red sequence and green valley galaxies. They concluded that accretion onto the central supermassive black hole may continue after the star formation has ceased, and thus if AGN feedback is responsible for the cessation of star formation in these galaxies, then the method by which this occurs does not completely remove gas from the nuclear regions.

In order to accurately compare theoretical predictions to conclusions based on observational studies, it is essential that we understand how the presence of an AGN may bias the colour and morphology measurements of its host galaxy. Morphologically, an unobscured AGN can be approximated as a point or as a highly concentrated object. Thus morphology measurements would be expected to experience a bias toward more bulge-like or spheroidal systems; the significance of this bias should depend strongly upon the balance between AGN and galactic light contributions. If the bias is significant and common, then we would need to reconsider the current classification of most AGN host galaxies as elliptical or early-type (e.g., Grogin et al.\ 2003, 2005; S\'{a}nchez et al.\ 2004; Pierce et al.\ 2007), with occasional morphological evidence of recent interactions (e.g., Bennert et al.\ 2008; Urrutia, Lacy \& Becker 2008).

One plausible explanation for the elliptical classification is that AGNs tend to be observed in galaxies that have recently undergone a merger. However, if the morphology measurements are significantly biased, then many of the apparently post-merger systems hosting AGNs may in fact be disc-dominated, necessitating an alternate explanation for the black hole growth. A potential bias toward elliptical morphologies may also cause an underestimate of the merger rate, as determined by morphology measurements such as $G$-$M_{20}$ (Lotz et al.\ 2008a). Thus, the proper classification of AGN host galaxy morphologies offers important clues as to when and how significant black hole growth is activated.

In a recent study of AGN host galaxy morphologies, Gabor et al.\ (2009) accounted for the presence of a point source component in the following two ways: (1) using GALFIT (Peng et al.\ 2002) to measure two-dimensional surface brightness profiles, Gabor et al.\ (2009) utilized the option to include a point spread function (PSF) as a model for the nuclear point source in the fitting procedure; (2) they also measured the concentration and asymmetry of the galaxies both before and after subtracting a modeled point source that had been fitted to the AGN. They found that both concentration and asymmetry measurements are affected by the presence (or subtraction) of a point source.

For the current study, we do not attempt to model a point source in the galaxies analyzed, but instead focus on understanding the effect that the presence of a point source may have on common morphology measurements. In AGN host galaxies, it is often unclear what fraction of the light emanates from the AGN and how much originates in star formation regions. A model of the nuclear light may thus overestimate the AGN contribution, and morphology measurements (such as $G$ and $M_{20}$) of the residual from a subsequent subtraction may experience a different set of biases than those caused by the actual point source. An investigation of such effects is beyond the scope of this paper.

Kim et al.\ (2008) emphasized the importance of using an appropriate PSF for studies similar to that described by Gabor et al.\ (2009), and they found that PSF characteristics can vary strongly with time, location on a camera detector, and level of pixel sampling. Mismatches between the PSF appropriate for a particular AGN and the PSF(s) determined for a set of observations may cause inaccurate measurements of the host galaxy and/or AGN fluxes. However, in \S~\ref{expt_1:visual_agn}, we find that the choice of PSF is negligible for the present study.

Measured colours may also be biased by light contributed by an AGN. Because optical and UV light from an AGN is typically much bluer than the light from its host galaxy (e.g., Kinney et al.\ 1996), a significant contribution from the AGN to the observed light may cause an intrinsically red AGN host galaxy to appear blue or `green', complicating or invalidating the use of colour to describe the host galaxy's star formation history. S\'{a}nchez et al.\ (2004) measured the colours and morphologies of 15 AGN host galaxies at $z \sim 1$, after first subtracting a PSF individually determined for each AGN. Upon comparing the resulting nuclear colours to the host galaxy colours, they found no adverse affects from the PSF subtraction. Though a few of their AGN host galaxies exhibit red colours, most show colours that are bluer than typical early-type galaxies, indicative of young stellar populations and consistent with the results from Schawinski et al.\ (2007) described above.

The current paper investigates the biases caused by an AGN on five morphology measurements and two colour measurements. Four of the five morphology measurements considered in the current work were initially calibrated without accounting for possible effects due to the presence of active nuclei (Lotz, Primack \& Madau 2004). The fifth measurement technique includes an option to account for the presence of a point source (Peng et al.\ 2002). AGN contamination of UV-optical colours was recently tested by Kauffmann et al.\ (2007), who reported that the amount of AGN light potentially contaminating measured UV-optical colours should be minimal for low-redshift ($z < 0.07$), optical spectra-selected AGNs. Seeking to more fully understand the effect of higher-luminosity and higher-redshift ($z \sim 0.4$) AGNs on morphology and colour measurements, this paper presents an examination of the extent to which an AGN similar to a quasar or a Type-1 Seyfert galaxy may affect measurements of the Gini coefficient $G$, the second order moment of the brightest 20 per cent of the galaxy's flux $M_{20}$, concentration $C$, asymmetry $A$, the \sersic\/ index $n$, and optical ($U-B$) and UV-optical (NUV$-r$) colours.

We begin with a description of the two sets of data used for our experiments (\S~\ref{data}), followed by descriptions of the effects of an AGN on the measured morphology (\S~\ref{expt_1}) and colour (\S~\ref{expt_2}) of its host galaxy. The two sets of results are combined in \S~\ref{nuvr_morph}, where we present a potential track of an unobscured AGN in colour-morphology space, as a function of the fraction of $B$ band light contributed by the AGN. Finally, in \S~\ref{summary} we discuss and summarize our results. Throughout, \{$h$, $\Omega_{\Lambda}$, $\Omega_{M}$\} $=$ \{$0.7$, $0.7$, $0.3$\}, and AB magnitudes are used, unless otherwise noted.

%%%%%%%%%%%%%%%%%%%%%%%%%%%%%
\section{Data}\label{data}

\subsection{AGN host galaxy images}\label{data:images}
The present study uses data available from the All-wavelength Extended Groth Strip International Survey (AEGIS; Davis et al.\ 2007, and references therein). In particular, we use optical images taken with the {\it Hubble Space Telescope}/Advanced Camera for Surveys ({\it HST}/ACS), optical spectra taken with the DEIMOS spectrograph (Faber et al.\ 2003) as part of the DEEP2 Redshift Survey (Davis et al.\ 2003, 2007) and with the MMT Observatory (Coil et al. 2009), deep {\it Chandra X-ray Observatory} images (Laird et al.\ 2009), and Very Large Array (VLA) images taken at 1.4 GHz (Ivison et al.\ 2007). The galaxies used for our first experiment are drawn from a region of AEGIS for which all of these data are available.

We select galaxies that have spectroscopic redshifts $0.2 < z < 0.6$, so that all the analyses can be performed using {\it HST}/ACS $V$ band images, which approximately correspond to the rest-frame $B$ band at these redshifts. The {\it HST}/ACS $V$ band images analysed here were taken with the F606W filter, using single-orbit exposures and four dithers. The resulting images have a PSF of $\sim 0.1\arcsec$ and a scale of $0.03\arcsec/$pixel; see Lotz et al.\ (2008a) for details regarding the reduction of the {\it HST}/ACS images.

Galaxies that have X-ray luminosities $L_{\rm 2-10 \ keV} > 10^{42}$ erg s$^{-1}$, radio powers $P_{\rm 1.4 \ GHz} > 10^{1.2z+23.2}$ W Hz$^{-1}$, and/or have been spectrally identified as Seyfert galaxies, LINERs, or broad-line AGNs are excluded from the sample in order to control the effect of an AGN.\footnote{Pierce (2009) described in greater detail the identification of X-ray, radio, and optical spectra-selected AGNs in the AEGIS and GOODS-N regions.} We intentionally exclude AGNs from our sample because the current study aims to determine the effect that an AGN might have on colour and morphology measurements of its host galaxy, and including AGNs would severely complicate these efforts.

A sample of galaxies meeting these criteria is created from 29 of the 63 {\it HST}/ACS tiles, each containing between 12 and 34 eligible galaxies. This sample of 622 galaxies is later refined by imposing criteria designed to assure the reliability of the individual morphology measurements (\S~\ref{expt_1:morph_measurements}). Because each of the individual and combined morphology measurements require different criteria, the final samples used for each morphology analysis differ slightly; additional descriptions of each sample are provided with the results.

\subsection{AGN spectral template}\label{data:agn_template}
Spectral templates are available for a variety of AGN types, including quasars (e.g., Francis et al.\ 1991), LINERs, and Seyfert galaxies (e.g., de Bruyn \& Sargent 1978). Compared to a quasar, lower luminosity and/or more heavily obscured AGNs are expected to less significantly bias the measured host galaxy colours, although some contamination will be unavoidable for all but the most heavily obscured AGNs. One goal of the current work is to determine the level of this bias in AGN host galaxies. Ideally, we would determine this for all AGN types, however, the intervening galactic matter present in obscured AGNs and the significant light contributions from the host galaxies of lower luminosity AGNs will contaminate the observed {\it AGN} light, biasing the measured colour of the AGN itself. Since this is the reverse of the problem that we want to address with the current study, we select a quasar template in order to avoid the potentially significant light contamination from the host galaxy. Although quasars are among the less common and more extreme AGN types at redshifts $z \sim 1$, the current work provides upper limits to the biases in the measured AGN host galaxy colours.

Francis et al.\ (1991) prepared a composite quasar spectral template based on more than 700 individual quasar and AGN spectra from the Large Bright Quasar Survey. The observations covered at least 3200-6800\AA\/ at a resolution of approximately 6\AA\/ and a S$/$N$ = 12$. After correcting for redshifts, between 200 and 500 AGNs contribute to the composite spectrum at rest-frame wavelengths $\lambda_{0} =$ 1200-4200\AA, while up to 200 contribute at $\lambda_{0} =$ 800-1200\AA\/ and $\lambda_{0} =$ 4200-6000\AA.

\begin{figure}
\begin{center}
\includegraphics[width=3.5in]{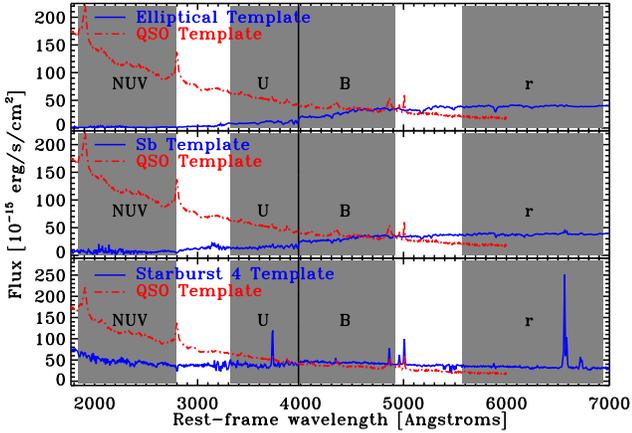}
\caption{Non-AGN (solid blue curves) and quasar (dot-dashed red curves) spectral templates. Gray shaded regions represent the effective widths of the four passbands indicated. Note that the $U$ and $B$ passbands do not overlap; the solid vertical line indicates where they split at $\lambda =$ 3980\AA.}
\label{fig:spectra}
\end{center}
\end{figure}

Figure~\ref{fig:spectra} shows the quasar composite spectrum along with the spectra of the three non-AGN galaxy templates described below; note that although the quasar template does not extend all the way across the $r$ band, we do not expect this to be problematic, due to the low contribution expected from the quasar in the $r$ band, relative to the $U$ and NUV bands. The most prominent feature of the quasar spectral template, as shown in Figure~\ref{fig:spectra}, is the broad MgII line at $\lambda =$ 2798\AA. H$\beta$ ($\lambda =$ 4861\AA) and [OIII] ($\lambda\lambda =$ 4959, 5007\AA) lines are also apparent. [Figure 2 of Francis et al.\ (1991) shows the quasar spectrum in greater detail.]

\subsection{Non-AGN galaxy spectral templates}\label{data:quiescent_template}
Kinney et al.\ (1996) presented spectral templates of several galaxy types -- including elliptical, Sb, and starburst galaxies -- based on observations using the {\it International Ultraviolet Explorer} ({\it IUE}), the Cerro Tololo Inter-American Observatory (CTIO), and the Kitt Peak National Observatory. Optical spectra observed at the CTIO were measured with slits $10\arcsec \times 20\arcsec$, designed to match the area of the {\it IUE\/} aperture. Table 1 of Kinney et al.\ (1996) lists properties of the individual galaxies observed, and their Table 2 specifies which galaxies are used for each template.

To create the spectral templates, Kinney et al.\ (1996) shifted each spectrum to its rest frame, corrected for Galactic extinction, and then averaged the UV and optical spectra within each galaxy morphology group (e.g., Sb). Optical spectra were averaged separately from UV spectra, but similar methods were employed for both averaging processes, and then the two averaged spectra were combined to produce the final templates (shown in Figure 2 of their paper); the templates used for the current paper are shown in Figure~\ref{fig:spectra}. Kinney et al.\ (1996) compared their non-AGN galaxy templates to results from Kennicutt (1992) and noted that their Sb template does not sufficiently represent young stellar populations in the disc. This suggests that colours derived from the Sb galaxy template may be too red. However, this should not significantly affect the work reported here.

%%%%%%%%%%%%%%%%%%%%%%%%%%%%%
\section{Experiment \#1: The Effect of an AGN on Morphology Measurements}\label{expt_1}

\subsection{Classifying galaxy morphologies}\label{expt_1:morph_measurements}
Galaxy morphologies depend upon the rest-frame wave-lengths of the images analysed. At any given redshift, the observed wavelengths that correspond to the rest-frame $B$ band are highly sensitive to recent star formation and young stellar populations; this distinction is particularly relevant for detection of recent or on-going galaxy interactions. Redder bands correspond to older stellar populations and may miss indications of recent activity, while bluer bands (e.g., NUV) will miss stellar populations that are no longer dominated by the youngest stars (e.g., O and B). At the redshifts used for the analyses described here ($0.2 < z < 0.6$), rest-frame $B$ band light approximately corresponds to the observed {\it HST}/ACS $V$ band images.

Code written by J.\ Lotz (e.g., Lotz et al.\ 2004) simultaneously measures $A$, $C$, $G$, and $M_{20}$, along with several additional quantities, from optical galaxy images. Lotz et al.\ (2004, 2008a) described a morphological classification system, using $G$ and $M_{20}$, that is particularly sensitive to interacting galaxies and is reliable to redshifts $z \sim 1$. In addition, Lotz et al.\ (2008b) have shown that $A$ and $G$ may be more effective than $C$ and $A$ at separating interacting galaxies from non-interacting galaxies.

Concentration and asymmetry (Conselice 2003) have been widely used by many authors to describe AGN host galaxies (e.g., Grogin et al.\ 2003, 2005); Pierce et al.\ (2007) instead used $G$ and $M_{20}$ for the same task. The light profiles of AEGIS, GOODS, GEMS, and COSMOS galaxies ($\sim$ 490,000 sources) have been measured by Griffith et al.\ (in preparation), using the GALAPAGOS\footnote{http://astro-staff.uibk.ac.at/$\sim$m.barden/galapagos/} code written by M. Barden (Barden et al.\ 2010), which implements the GALFIT program (Peng et al.\ 2002) to determine \sersic\/ indices.

\subsubsection{Asymmetry}\label{expt_1:asym}
Measuring the asymmetry $A$ of a galaxy is a three-step process, beginning with a rotation of the galaxy image (which includes all galaxy pixels within 1.5 r$_{\rm ell}$, where r$_{\rm ell}$ is the elliptical Petrosian radius; Petrosian 1976) by 180$^{\circ}$ about its central axis; in practice, the central axis is determined by minimizing $A$ (Lotz et al.\ 2004). Next the rotated image is subtracted from the original image, and finally, the sum of the absolute value of the residual flux is compared to the flux of the original galaxy (Conselice et al.\ 2000; Conselice 2003). This process is summarized by the following formula (Conselice et al.\ 2008):
\begin{equation}
A = min \left ( \frac{\Sigma |I_{0} - I_{180}|}{\Sigma |I_{0}|} \right ) - min \left ( \frac{\Sigma |B_{0} - B_{180}|}{\Sigma |B_{0}|} \right )
\end{equation}
where the subscripts refer to the rotation of the image, and $I$ and $B$ refer to the original image and the background, respectively. Highly symmetrical galaxies, such as undisturbed ellipticals or discs, leave minimal residuals, and $A$ is small; the significant residuals of disturbed and irregular galaxies correspond to higher values of $A$. Galaxies with $0 < A < 1$ are typical.

\subsubsection{Concentration}\label{expt_1:conc}
Concentration $C$ compares the radius enclosing a large fraction of a galaxy's flux to the radius enclosing a smaller fraction of its flux. For this work, we used the following definition (Conselice 2003):
\begin{equation}
C = 5 \times {\rm log} (r_{80}/r_{20}),
\end{equation}
where $r_{\rm n}$ represents the radius enclosing n per cent of the flux within 1.5 r$_{\rm ell}$ of the galaxy's centre. High values of $C$ correspond to highly concentrated systems, such as elliptical galaxies, and the value of $C$ decreases for more diffuse galaxies, generally including disc galaxies and many interacting systems. Concentrations $2 < C < 5$ are typical values for observed galaxies.

\subsubsection{The Gini coefficient}\label{expt_1:gini}
Abraham et al.\ (2003) were the first to use $G$ to describe the distribution of light among the pixels associated with a galaxy, though it was initially introduced as an economic statistic used to characterize the distribution of wealth in a society (Gini 1912). Lotz et al.\ (2004, 2008a) describe in detail the measurement of $G$ and $M_{20}$; here we provide brief summaries.

If $n$ is the number of pixels assigned to a galaxy and $f_{i}$ represents the flux from pixel $i$ (ordered such that $f_{i}$ increases with the pixel index), then we can use the methods described by Lotz et al.\ (2004, 2008a) and measure $G$ using
\begin{equation}
G = \frac{1}{|\overline{f}|n(n-1)} \sum_{i}^{n} (2i-n-1)|f_{i}|.
\end{equation}
High values of $G$ correspond to galaxies in which the light is distributed among a small fraction of the associated pixels, such as a galaxy featuring a single bright nucleus or a galaxy that features multiple bright nuclei; this latter characteristic contributes to the sensitivity to interacting galaxies associated with the $G$-$M_{20}$ and $G$-$A$ classifications. Low values of $G$ correspond to a homogeneous distribution of light among the galaxy's pixels, such as would be found for a smooth disc galaxy. Gini measurements of $0.3 < G < 0.7$ are common for observed galaxies.

\subsubsection{$M_{20}$}\label{expt_1:m20}
The value of $M_{20}$ depends on the spatial distribution of the light in a galaxy, relative to the centre of the galaxy. Lotz et al.\ (2004) defined $M_{20}$ as follows:
\begin{equation}
M_{20} \equiv {\rm log}_{10} \left ( \frac{\sum_{i}M_{i}}{M_{\rm tot}} \right ), \ {\rm while} \ \sum_{i} f_{i} < 0.2 f_{\rm tot},
\end{equation}
where
\begin{equation}
M_{\rm tot} = \sum_{i}^{n} M_{i} = \sum_{i}^{n} f_{i}[(x_{i} - x_{c})^{2} + (y_{i} - y_{c})^{2}].
\end{equation}
Here, the fluxes $f_{i}$ are arranged in order of {\it decreasing} flux, and ($x_{c}$, $y_{c}$) is the galaxy centre, as determined by minimizing M$_{\rm tot}$. The presence of multiple bright nuclei that are spatially separated from a galaxy's centre increases the value of $M_{20}$ toward $-1.0$. However, if the brightest 20 per cent of the galaxy's light is positioned very close to the galactic centre, as would be the case for an undisturbed elliptical galaxy, the value of $M_{20}$ approaches $-2.5$ or $-3.0$. Typical observed galaxies have $-0.5 > M_{20} > -2.5$.

\subsubsection{Reliability criteria for $G$, $M_{20}$, $C$, and $A$}\label{expt_1:morph_criteria}
$G$, $M_{20}$, and $C$ are considered reliable for galaxies that have $\langle S/N \rangle$ per pixel $\ge 2.5$, elliptical Petrosian radii r$_{\rm ell} \ge 0.3\arcsec$, and contiguous segmentation maps (as determined by J.\ Lotz's code; Lotz et al.\ 2004, 2008a). Reliability criteria for measurements of $A$ differ only in that the $\langle S/N \rangle$ per pixel must exceed 4.0, instead of 2.5, a lower limit similar to that suggested by Lotz et al.\ (2004).

\subsubsection{\sersic\/ profiles}\label{expt_1:sersic}
The \sersic\/ profile of a galaxy is given by
\begin{equation}
\Sigma(r) = \Sigma_{e} \ \rm{exp} \left \{ -\kappa \left [ \left ( \frac{r}{r_{e}} \right )^{1/n}-1 \right ] \right \},
\end{equation}
where $n$ is the \sersic\/ index of the galaxy; $n = 1$ represents a purely exponential light profile (a disc galaxy), and $n = 4$ represents a purely de Vaucouleurs light profile (an elliptical galaxy).

Following H$\ddot{\rm a}$ussler et al.\ (2007) and Griffith et al.\ (in preparation), we use GALFIT Version 2.1c (Peng et al.\ 2002) and the {\it HST}/ACS $V$ band PSF to measure the \sersic\/ indices of the original and simulated galaxies in our sample. For the initial guesses allowed by GALFIT, we use the following values: $n = 2.5$, $b/a = 1-$\verb|ELLIPTICITY|, $r_{e} = 0.162 \times $\verb|FLUX_RADIUS|$^{1.87}$, and mag$=$\verb|MAG_BEST|; \verb|ELLIPTICITY|, \verb|FLUX_RADIUS|, and \verb|MAG_BEST| are determined by SExtractor\footnote{ftp://ftp.iap.fr/pub/from\_users/bertin/sextractor/} (Bertin \& Arnouts 1996).

Reliable GALFIT results are selected by requiring relative uncertainties of less than 15 per cent of the effective radii and the \sersic\/ indices ($\Delta r_{e} \le 0.15 \times r_{e}$; $\Delta n \le 0.15 \times n$), GALFIT magnitudes and SExtractor magnitudes agreeing to within 1.0 mag ($|m_{\rm GALFIT} - m_{\rm SExtractor}| \le 1.0$), and \sersic\/ indices $0.2 < n < 1.5$ or $1.5 < n < 8$. It is also required that GALFIT converges on a value and does not crash while analyzing the galaxy (Griffith et al., in preparation).

We exclude from our analyses galaxies to which GALFIT has assigned a \sersic\/ index $n = 1.5$ because we find that an increase of the fraction of light contributed by a simulated AGN (particularly for AGN fractions $>30$ per cent) corresponds to an increase in the number of galaxies having \sersic\/ indices $n=8$ and $n=1.5$. The former is expected, as the PSF used to simulate the AGNs has a \sersic\/ index $n=8$, but the latter is unexpected. Visual examination of the excluded galaxies reveals unusually prominent nuclear regions, such that we do not expect to {\it observe} similar galaxies. Therefore, excluding such galaxies does not limit the applicability of our analyses to observed AGN host galaxies.

A plausible explanation is that the PSF component of these systems dominates the fitting process of GALFIT, and thus the galaxy profile is not properly measured; whether the \sersic\/ index is assigned $n=8$ or $n=1.5$ depends on the specific characteristics of the galaxy and the profile fitting (C.\ Y.\ Peng, private communication). Although a full analysis of the cause behind this is beyond the scope of this paper, the convergence at $n=1.5$ disappears when we include PSF fitting while running GALFIT on a subset ($\sim 40$) of the simulated AGN host galaxies. Thus the convergence at $n=1.5$ may result in part from not fitting for a point source.

\subsubsection{Definitions of morphology classifications}\label{expt_1:morph_class}
Based on calibrations using more than 200 low-redshift galaxies, Lotz et al.\ (2004) found that the sensitivity of both $G$ and $M_{20}$ to multiple nuclei makes the $G$-$M_{20}$ classification technique particularly well suited for the identification of interacting galaxies. They also defined three morphology classifications for the non-interacting galaxies. Lotz et al.\ (2008a) refined the interacting and non-interacting classifications for high-redshift galaxies and additionally suggested classifications based on combining $G$ with $A$, which is also commonly combined with $C$ (Conselice 2003). Table 1 summarizes the morphology classification definitions adopted for this paper; the $G$-$M_{20}$ and $C$-$A$ definitions exactly follow Lotz et al.\ (2008a) and Conselice (2003), respectively, while the $G$-$A$ definitions combine the definitions suggested by the two references.

Morphology classifications based on the \sersic\/ indices are also summarized in Table 1, and are based on work by, e.g., Blanton et al.\ (2003), Trujillo et al.\ (2007), and Bell (2008). The `bulge$+$disc' classification indicates a \sersic\/ index between the disc-dominated and bulge-dominated classifications.

These classifications do not always suggest the same morphology for a given galaxy. For example, a single galaxy may be classified as interacting by $G$-$M_{20}$ and $G$-$A$, early-type by $C$-$A$, and bulge$+$disc by the \sersic\/ index [e.g., Figure~\ref{fig:gal_images}, Row (a)]. Visual morphology classification is therefore still an important check on the morphologies based on the location in $G$/$M_{20}$/$CAS$/\sersic\/ index space. However, the focus of the current study is an investigation of the effect of an AGN on the {\it automated} morphology measurements, so we adopt the morphology classifications determined by the automated methods.

\begin{table}
\centering
\begin{minipage}{85mm}
\caption{Morphology classification definitions. `Merger' is a sub-category of `Interacting' that specifically refers to the galaxy coalescence stage of an interaction.\label{table:morph_definitions}}
\begin{tabular}{@{}ll@{}}
  Morphology & Definition \\
\hline
\multicolumn{2}{c}{$G$ - $M_{20}$} \\
\hline
Interacting & $G > -0.14 \ M_{20} + 0.33$ \\
E/S0/Sa     & $G < -0.14 \ M_{20} + 0.33$ \& $G > 0.14 \ M_{20} + 0.80$ \\
Sb-Ir       & $G < -0.14 \ M_{20} + 0.33$ \& $G < 0.14 \ M_{20} + 0.80$ \\
\hline
\multicolumn{2}{c}{$G$ - $A$} \\
\hline
Interacting     & $G > -0.4 \ A + 0.66$ or $A > 0.35$ \\
Non-interacting & $G < -0.4 \ A + 0.66$ \& $A < 0.35$ \\
\hline
\multicolumn{2}{c}{$C$ - $A$} \\
\hline
Merger     & $A > 0.35$ \\
Early-type & $A < 0.35$ \& $C > 3.5$ \\
Late-type  & $A < 0.35$ \& $C < 3.5$ \\
\hline
\multicolumn{2}{c}{\sersic\/ Index $n$} \\
\hline
Disc         & $0.2 < n < 1.5$ \\
Bulge$+$disc & $1.5 < n < 2.5$ \\
Bulge        & $2.5 < n < 8.0$ \\
\hline
\end{tabular}
\end{minipage}
\end{table}

\subsection{Simulation of an optically visible AGN}\label{expt_1:visual_agn}
We use the Tiny Tim software (Krist 1993), which was specifically designed to model the PSFs for observations taken with various {\it HST} cameras, to determine appropriate $V$- and $I$-band PSFs for the AEGIS {\it HST}/ACS observations. For the current work, the PSF is contained within a square array of 121 pixels $\times$ 121 pixels, designed to complement the size of postage stamps created from the original galaxy images. After selecting the 622 non-AGN galaxies that constitute our initial sample, the presence of an AGN is simulated in each galaxy by adding the {\it HST}/ACS PSF to each galaxy's $V$ band image. The same PSF is provided to GALFIT for its determination of the \sersic\/ profiles. However, we do not use GALFIT to explicitly fit for the presence of the PSF, rather it is used for image convolution to account for blurring (Peng et al.\ 2002).

As described by, e.g., Kim et al.\ (2008), the choice of PSF can significantly affect galaxy morphology measurements. Although it is not the goal of the current paper to investigate PSF fitting, we do want to understand how the choice of PSF may affect our results. Therefore, we also run GALFIT providing a PSF based on {\it HST}/ACS observations and find that our results do not significantly depend upon the input PSF.

The {\it HST}/ACS images used for the current study have been drizzled, but the PSF has not. Because $G$ and $M_{20}$ specifically analyse the light within pixels, we want to know how this may affect our morphological analyses. To do this, we select a star from one of the drizzled ACS images used for the current work that is approximately the same size as the PSF. After running the morphology code on both the star and the Tiny Tim PSF, we find good agreement between the two sets of measurements; the differences between the morphology measurements are $|\Delta G| = |\Delta M_{20}| = 0.01$.

Taking the SExtractor-derived value \verb|MAG_BEST| as an approximation of the apparent magnitude of each galaxy, the $V$ band fluxes of the initial galaxies are calculated as follows: f$_{\rm gal} = 10 ^{-0.4 \times (\verb|MAG_BEST|-26.486)}$, where f$_{\rm gal}$ has units of [counts$/$s]. We then determine the scaling factors necessary to cause the PSF to contribute a fraction $\%{\rm AGN}$ of the total $V$ band flux observed from the resulting galaxy/AGN system, where f$_{\rm AGN} =$ f$_{\rm gal} \times [\%_{\rm AGN} /(1 - \%_{\rm AGN}$)]. Values of $\%_{\rm AGN}$ range from 0 per cent (the initial galaxy) to 50 per cent, and increase by 5 per cent increments ($\%_{\rm AGN} =$ 0 per cent, 5 per cent, 10 per cent, 15 per cent, etc.).

When a galaxy's natural position of maximum intensity (as defined by SExtractor's \verb|X_PEAK| and \verb|Y_PEAK|; Bertin \& Arnouts 1996) is offset from the galaxy's geometrical centre, increasing the brightness of the geometrical centre may result in a galaxy image containing two bright nuclei. Although this could allow insightful tests of the morphology measurements, the current goal is to increase the intensity of the natural nucleus. Therefore, when the scaled PSFs are added to the galaxy images, the centre of the PSF array is positioned to coincide with (\verb|X_PEAK|, \verb|Y_PEAK|) from the SExtractor measurements of the original galaxy images.

\begin{figure*}
\begin{center}
\includegraphics[width=7in]{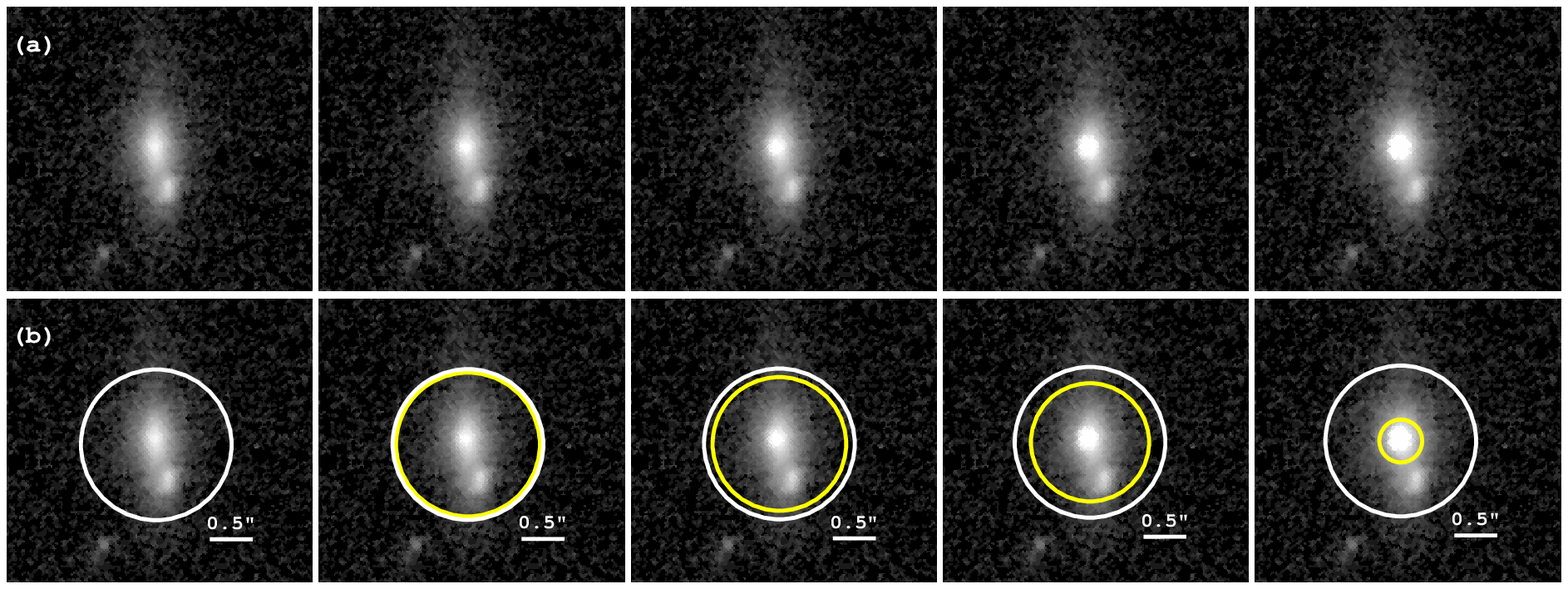}
\includegraphics[width=7in]{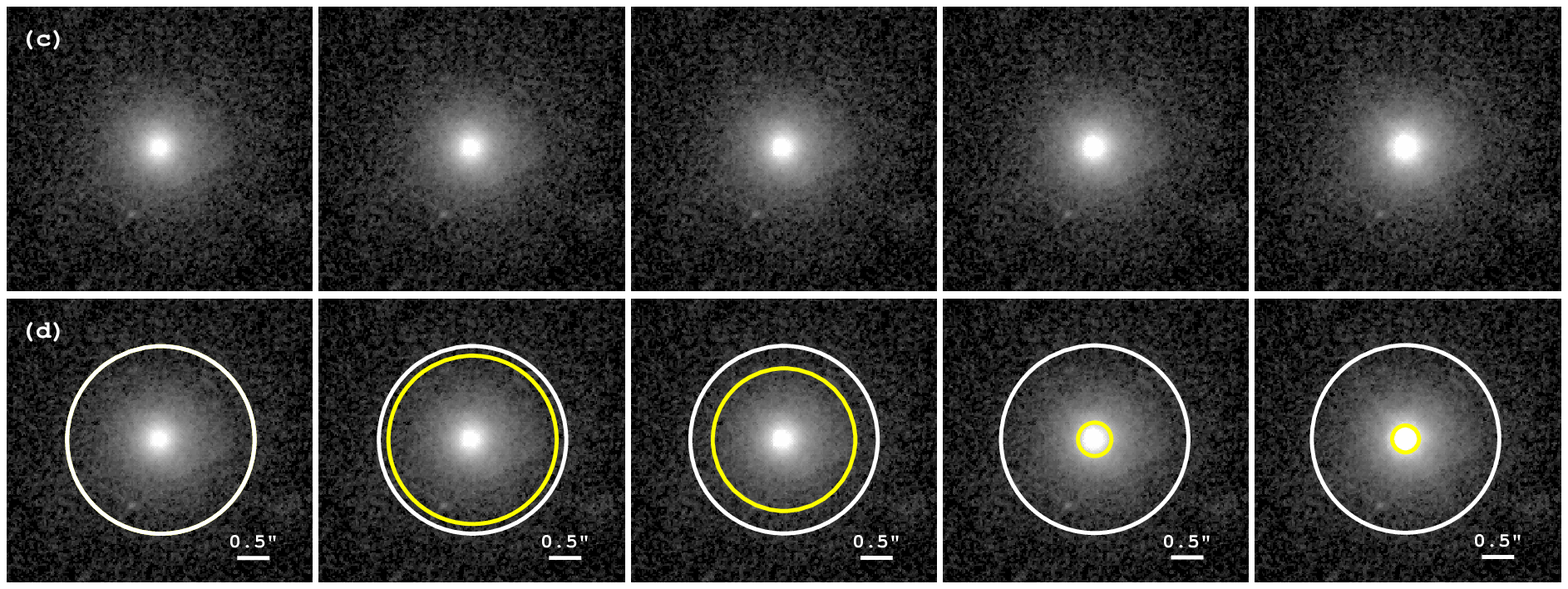}
\includegraphics[width=7in]{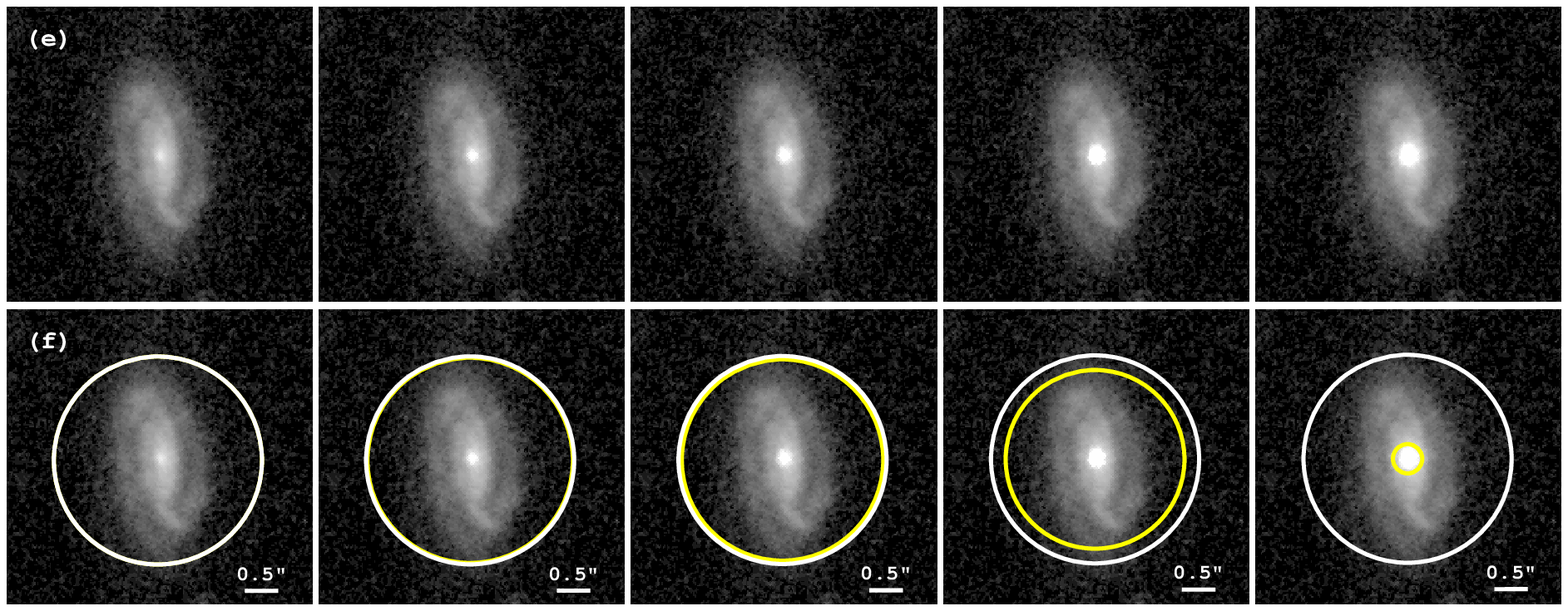}
\caption{Three examples of non-AGN galaxies in which we simulate an AGN represented by a point spread function. For each set, the fraction of $B$ band light contributed by the AGN (0, 5, 10, 20, and 30 per cent) increases to the right. The images are shown on a log scale ranging from 0-40, with a contrast of 2.02 and a bias of 0.247. {\it Row} ({\it a}): Galaxy 13009807: $z = 0.335$; $G$-$M_{20}$ : interacting; $G$-$A$: interacting; $C$-$A$: early-type; \sersic\/ profile: bulge$+$disc. {\it Row} ({\it b}): Same galaxy as row (a); the radius of the white circles is the Petrosian radius measured for the initial galaxy, and the radii of the yellow circles are the Petrosian radii measured for the galaxies with PSFs added; the bar is 0.5 $\arcsec$ in length. {\it Row} ({\it c}): Galaxy 13018030: $z = 0.450$; $G$-$M_{20}$ : E/S0/Sa; $G$-$A$: non-interacting; $C$-$A$: early-type; \sersic\/ profile: bulge-dominated. {\it Row} ({\it d}): Same galaxy as row (c), with circles as described for row (b). {\it Row} ({\it e}): Galaxy 13017884: $z = 0.385$; $G$-$M_{20}$: Sb-Ir; $G$-$A$: non-interacting; $C$-$A$: late-type; \sersic\/ profile: disc-dominated. {\it Row} ({\it f}): Same galaxy as row (e), with circles as described for row (b).}
\label{fig:gal_images}
\end{center}
\end{figure*}

Figure~\ref{fig:gal_images} shows three of the galaxies in which an AGN has been simulated. The original {\it HST}/ACS $V$ band image of each galaxy is shown, followed by $V$ band images of the galaxy for which 5, 10, 20, and 30 per cent of the observed $V$ band light is contributed by an AGN, with the AGN fraction increasing to the right. Rows (a), (c), and (e) show the original and altered galaxy images. For rows (b), (d), and (f), we overplot the images from the preceding rows with circles indicating the original and current Petrosian radii. For all of the images, the light is shown on a log scale, using the same minimum, maximum, contrast, and bias (0, 40, 2.02, and 0.247, respectively). These three galaxies initially meet the reliability criteria (\S~\ref{expt_1:morph_criteria}) for all of the morphology measurements described above, and the figure caption provides the $G$-$M_{20}$, $G$-$A$, $C$-$A$, and \sersic\/ profile morphology classifications based on the original galaxies.

Based on the visibility and distinct outlines of the simulated AGNs, the AGN shown in row (a) is visually identifiable at an AGN fraction of 20 per cent, the AGN shown in row (e) is visually identifiable at an AGN fraction of only 10 per cent, and the AGN shown in row (c) appears visually identifiable at an AGN fraction of 20-30 per cent. The presence of a bright nucleus does not seem out of place for the elliptical galaxy shown in row (c), complicating the visual distinction between a star forming bulge and an AGN for such galaxies. As indicated by the yellow (inner) circles in Figure~\ref{fig:gal_images}, the Petrosian radii decrease significantly for AGN fractions exceeding 20-30 per cent. Thus, the size of the Petrosian radius correlates with the visibility of the AGN; as the former decreases, the latter increases.

\subsection{Results}\label{expt_1:results}
In this section, we describe the effects of an AGN on the elliptical Petrosian radius and individual (e.g., $G$, $C$) and paired (e.g., $G$-$M_{20}$) morphology measurements. In the figures, the symbol shapes and colours indicate the initial morphology classifications of the galaxies represented. For example, when presenting the results for $G$ and $M_{20}$, filled red circles represent systems that are initially classified by $G$-$M_{20}$ as E/S0/Sa galaxies. The number of objects represented in each figure is indicated on the individual plots; it is found to rapidly decline for $\%_{\rm AGN} >$ 20 per cent, most often due to a shrinking r$_{\rm ell}$.

\subsubsection{Individual morphology measurements}\label{expt_1:individual_morphs}
\begin{figure}
\begin{center}
\includegraphics[width=3.5in]{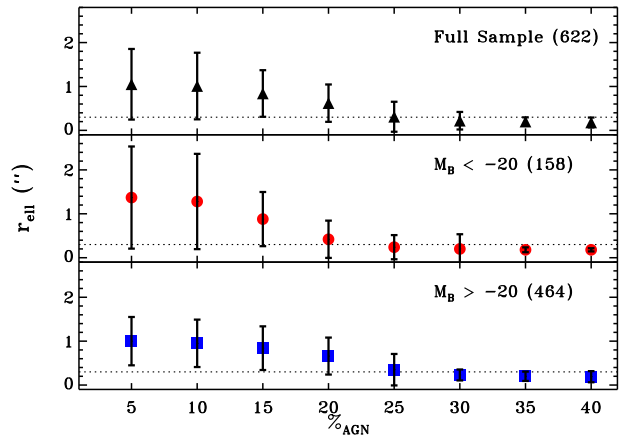}
\caption{Elliptical Petrosian radius r$_{\rm ell}$ as a function of $\%_{\rm AGN}$. Symbols represent median radii for the full sample of galaxies (top), bright galaxies ($M_{B} < -20$; middle) and faint galaxies ($M_{B} > -20$; bottom). Black error bars indicate the standard deviations for each $\%_{\rm AGN}$. The dotted line indicates the minimum required radius, r$_{\rm ell}$ = 0.3\arcsec, for the morphology measurements used in the current study.}
\label{fig:rell}
\end{center}
\end{figure}

The reliability criteria of $A$, $C$, $G$, and $M_{20}$ measurements require that a galaxy's elliptical Petrosian radius, r$_{\rm ell}$, exceed 0.3\arcsec (\S~\ref{expt_1:morph_criteria}). However, as the light observed from a galaxy/AGN system is increasingly dominated by the AGN, the measured r$_{\rm ell}$ decreases toward the size of the AGN. Figure~\ref{fig:rell} shows the median elliptical Petrosian radii r$_{\rm ell}$ for the entire galaxy sample (top panel), bright galaxies ($M_{B} < -20$; middle panel), and faint galaxies ($M_{B} > -20$; bottom panel). At AGN fractions of 25 per cent or more, the median radii are similar for all three samples, while at AGN fractions of 20 per cent or less, the median radii of the brighter galaxies are slightly larger than the median radii of the full and faint samples.

For AGN fractions greater than 20 per cent, the median Petrosian radii of galaxies in the full sample are smaller than the required 0.3\arcsec (dotted line in Figure~\ref{fig:rell}). This causes a significant decrease in the number of galaxies for which the morphology measurements studied here are reliable, which has two serious implications. (1) The galaxies that continue to provide reliable morphologies at the higher AGN fractions are not representative of the general population of AGN host galaxies; these are systems for which the extended regions are bright enough to successfully compete with the increasingly bright AGN light in the nuclear regions. The trends shown in the next several figures should be considered with this in mind. (2) Figure~\ref{fig:rell} suggests that most of the AGNs in systems for which we {\it do} have reliable morphology measurements contribute little more than 20 per cent of the observed optical light. This conclusion is supported by results from Pierce et al.\ (2007), who studied two samples of AEGIS AGN host galaxies at redshifts $0.2 < z < 1.2$ and measured reliable morphologies ($G$ and $M_{20}$ ) for at least 80 per cent of their AGN samples. This suggests that the majority of AGNs at $z \sim 1$ contribute less than 20 per cent of the optical light observed from the galaxy/AGN systems.

\begin{figure}
\begin{center}
\includegraphics[width=3.5in]{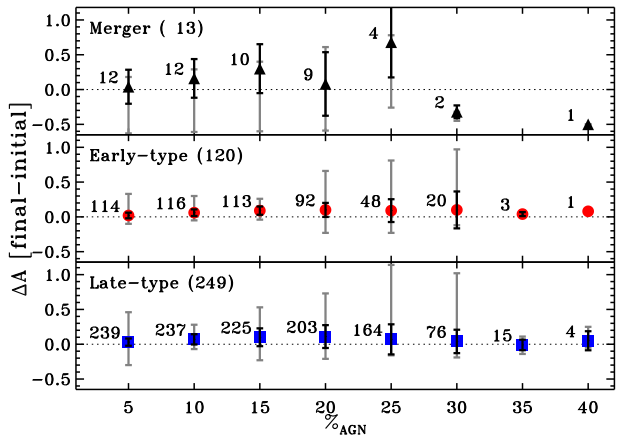}
\caption{$\Delta A$ as a function of $\%_{\rm AGN}$. The symbols indicate the median changes $\Delta A = A_{\rm final} - A_{\rm initial}$, black error bars indicate the standard deviations for each $\%_{\rm AGN}$, and gray error bars indicate the maximum and minimum changes at each $\%_{\rm AGN}$. Symbol shapes and colours indicate whether the original galaxies are classified by $C$-$A$ as a merger (black triangles), an early-type (red circles), or a late-type (blue squares). The numbers in parentheses indicate the initial number of galaxies assigned to each morphology classification, and the number of galaxies having reliable $A$ measurements at each $\%_{\rm AGN}$ is indicated next to the corresponding symbol.}
\label{fig:asym}
\end{center}
\end{figure}

Unless the simulated AGN is positioned at a location significantly different from the geometrical centre of a galaxy, the change in $A$ as $\%_{\rm AGN}$ increases is minimal, as shown by Figure~\ref{fig:asym}, a plot of the median changes $\Delta A = A_{\rm final} - A_{\rm initial}$ as a function of the AGN fraction. Galaxies initially classified as mergers do appear increasingly asymmetric as $\%_{\rm AGN}$ increases to 25 per cent; the decrease for higher AGN fractions should indicate that the morphology code is focusing on the PSF, instead of the full galaxy.

The focus of the morphology code on the PSF is also evident by the decreasing number of galaxy/AGN systems that provide reliable morphology measurements. At an AGN fraction of 10 per cent, 31 per cent (191/622) of the systems do not have reliable asymmetry measurements; 14 per cent (27/191) of these drop-outs are measured to have elliptical Petrosian radii r$_{\rm ell} < 0.3\arcsec$. For an AGN fraction of 20 per cent, we find a slightly higher overall drop-out rate (33 per cent; 207/622), but a significant fraction (72 per cent; 150/207) of these are are excluded because r$_{\rm ell} < 0.3\arcsec$. This pattern continues as the AGN fraction increases.

\begin{figure}
\begin{center}
\includegraphics[width=3.5in]{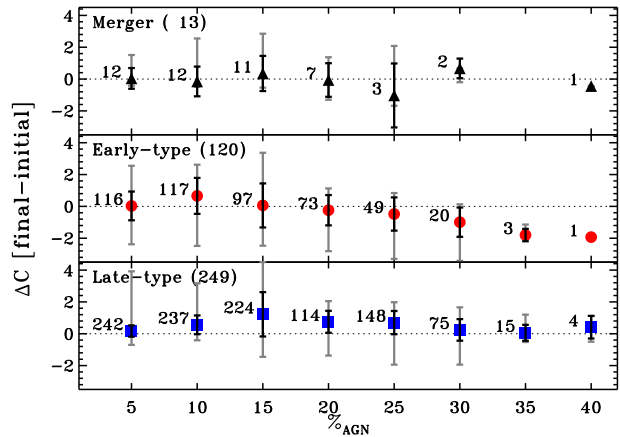}
\caption{$\Delta C$ as a function of $\%_{\rm AGN}$. The symbols indicate the median changes $\Delta C = C_{\rm final} - C_{\rm initial}$, and the error bars and symbols are as in Figure~\ref{fig:asym}. The numbers in parentheses indicate the initial number of galaxies assigned to each morphology classification, and the number of galaxies having reliable $C$ measurements at each $\%_{\rm AGN}$ is indicated next to the corresponding symbol. The measurements either increase as expected or are minimally affected for $\%_{\rm AGN} <$ 20 per cent; above this the measurements become unreliable.}
\label{fig:conc}
\end{center}
\end{figure}

Figure~\ref{fig:conc} shows that increasing the AGN fraction generally causes an increase or a minimal change in $C$ for $\%_{\rm AGN} <$ 20 per cent. As defined in \S~\ref{expt_1:conc}, $C$ is a ratio of the radii enclosing 20 per cent and 80 per cent of a galaxy's light, thus concentration measurements should only be reliable for $\%_{\rm AGN} <$ 20 per cent, otherwise the inner radius simply encloses the PSF that has been added to the initial system, and the outer radius approaches the same location as the $\%_{\rm AGN}$ increases. For certain host galaxies, the sensitivity of $C$ to the defined galaxy centre may provide an alternate explanation for negative values of $\Delta C$. This would be particularly relevant for AGNs offset from the galactic centre.

As for the $A$ measurements, the number of reliable meausurements of $C$ depends on the fraction of light contributed by the AGN. Twenty per cent (124/622) of the $C$ measurements are deemed unreliable at an AGN fraction of 10 per cent; we measure elliptical Petrosian radii r$_{\rm ell} < 0.3\arcsec$ for 22 per cent (27/124) of these. For an AGN fraction of 20 per cent, 56 per cent (348/622) of the systems do not provide reliable morphology measurements, and 43 per cent (150/348) of these are due to small elliptical Petrosian radii.

\begin{figure}
\begin{center}
\includegraphics[width=3.5in]{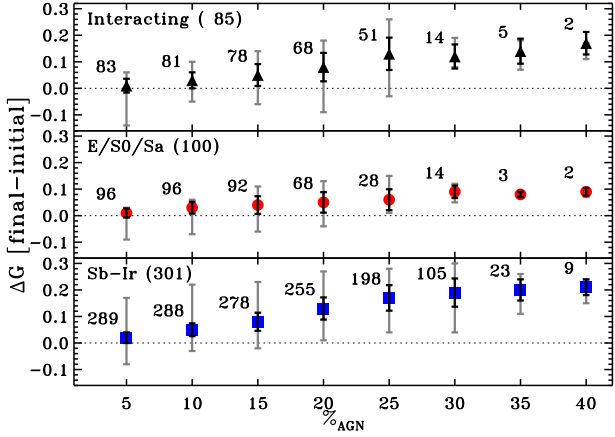}
\caption{$\Delta G$ as a function of $\%_{\rm AGN}$. The symbols indicate the median changes $\Delta G = G_{\rm final} - G_{\rm initial}$, and the error bars are as in Figure~\ref{fig:asym}. Symbol shapes and colours indicate whether the original galaxies are classified by $G$-$M_{20}$ as interacting (black triangles), E/S0/Sa (red circles), or Sb-Ir (blue squares). The numbers in parentheses indicate the initial number of galaxies assigned to each morphology classification, and the number of galaxies having reliable $G$ measurements at each $\%_{\rm AGN}$ is indicated next to the corresponding symbol. Sb-Ir galaxies are the most significantly affected by the presence of a bright AGN, and E/S0/Sa galaxies are the least affected; interacting galaxies provide an unexpected result with several examples of decreasing $G$, instead of the expected increase.}
\label{fig:gini}
\end{center}
\end{figure}

Figure~\ref{fig:gini} shows the median changes $\Delta G = G_{\rm final} - G_{\rm initial}$ for each of the initial morphology classifications. It is clear from this figure that Sb-Ir galaxies are the most significantly affected by the presence of a bright AGN, and E/S0/Sa galaxies are the least affected. Several galaxies, particularly those initially classified as interacting, exhibit a {\it decrease} in $G$ as the AGN fraction is increased, instead of the expected increase. The disturbed morphologies of some of these galaxies, combined with the added PSF, may result in an inconsistently defined boundary to the galaxy, thereby confusing measurements of $G$. However, $G$ may be more reliable than $C$ for high $\%_{\rm AGN}$, in part because $G$ is less sensitive to a defined galaxy centre.

The number of systems for which $G$ is reliable also depends on the AGN fraction. At an AGN fraction of 10 per cent, 22 per cent of the unreliable measurements are caused by r$_{\rm ell} < 0.3\arcsec$, 22 per cent have low $\langle$S/N$\rangle$ per pixel, and 80 per cent (99/124) of the unreliable systems are deemed non-contiguous (\S~\ref{expt_1:morph_criteria}); some systems fail multiple criteria, so the sum of the fractions exceeds 100 per cent. In contrast, at an AGN fraction of 20 per cent, elliptical Petrosian radii r$_{\rm ell} < 0.3\arcsec$ account for 88 per cent of the unreliable systems, low $\langle$S/N$\rangle$ per pixel account for 15 per cent, and non-contiguity accounts for 19 per cent (again, some systems fail multiple criteria). Thus, non-contiguous systems are most problematic at lower AGN fractions, while small elliptical Petrosian radii are most problematic at higher AGN fractions.

\begin{figure}
\begin{center}
\includegraphics[width=3.5in]{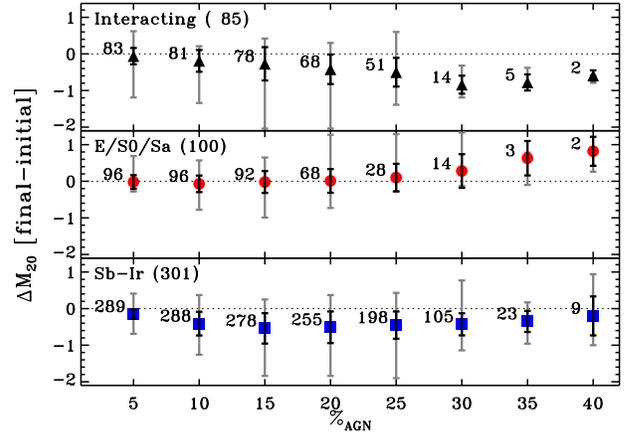}
\caption{$\Delta M_{20}$ as a function of $\%_{\rm AGN}$. The symbols indicate the median changes $\Delta M_{20} = M_{\rm 20,final} - M_{\rm 20,initial}$; error bars are as in Figure~\ref{fig:asym} and symbol shapes are as in Figure~\ref{fig:gini}. The numbers in parentheses indicate the initial number of galaxies assigned to each morphology classification, and the number of galaxies having reliable $M_{20}$ measurements at each $\%_{\rm AGN}$ is indicated next to the corresponding symbol. The median changes for E/S0/Sa galaxies are not as strong as the changes experienced by galaxies in the other two morphology classifications.}
\label{fig:m20}
\end{center}
\end{figure}

As $\%_{\rm AGN}$ increases toward 20 per cent, $M_{20}$ tends to decrease for Sb-Ir galaxies and interacting galaxies, as expected, while the measured values of $M_{20}$ for E/S0/Sa galaxies tend to remain unchanged. These effects are demonstrated in Figure~\ref{fig:m20}, a plot of the median changes $\Delta M_{20} = M_{\rm 20,final} - M_{\rm 20,initial}$. The reliability criteria for $M_{20}$ are identical to those for $G$, so the explanation for the decrease in the number of reliable measurements is the same as described in \S~\ref{expt_1:results}.

The unexpected increases in the $M_{20}$ measurements of E/S0/Sa galaxies and the large dispersions in $\Delta M_{20}$ for Sb-Ir and interacting galaxies at $\%_{\rm AGN} \sim$ 20 per cent can be at least partially understood by a review of the definition of this measurement (\S~\ref{expt_1:m20}). Of particular importance are the locations of the brightest 20 per cent of a galaxy's pixels. If an AGN contributes at least 20 per cent of the light from the system, many of the brightest pixels will be associated with the AGN, and thus restricted to a particular location. In general, this will significantly bias the measurement toward lower values of $M_{20}$ . On occasion, the galaxy may have a few pixels outside the vicinity of the AGN that are bright enough to compete with the AGN light, so the classification is not inherently doomed for bright AGNs, but it is very likely to be misleading.

The \sersic\/ index $n$ is a measure of a galaxy's light profile, an indication of whether the galaxy is better represented as a disc ($n = 1$) or a bulge ($n = 4$). Increasing the brightness of a few adjacent pixels would tend to increase the \sersic\/ index toward a bulge profile.

\begin{figure}
\begin{center}
\includegraphics[width=3.5in]{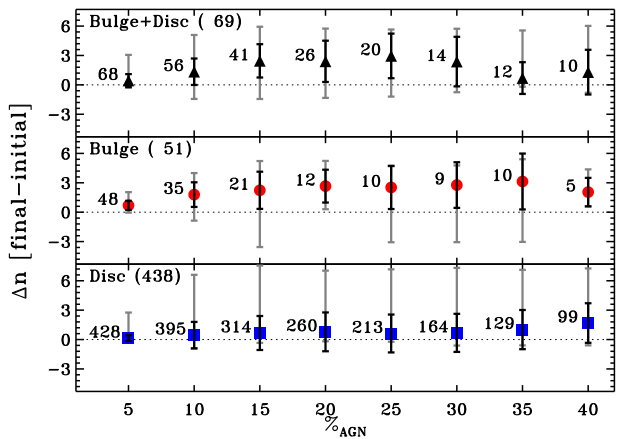}
\caption{$\Delta n$ as a function of $\%_{\rm AGN}$. Symbols indicate the median values of $\Delta n = n_{\rm final} - n_{\rm initial}$, and error bars are as in Figure~\ref{fig:asym}. Symbol shapes and colours indicate whether galaxies are initially classified as bulge-dominated (red circles), disc-dominated (blue squares), or bulge$+$disc (black triangles). The numbers in parentheses indicate the initial number of galaxies assigned to each morphology classification, and the number of galaxies having reliable \sersic\/ indices at each $\%_{\rm AGN}$ is indicated next to the corresponding symbol. The median changes experienced by initially disc-dominated systems are clearly less than the changes experienced by initially bulge$+$disc or bulge-dominated systems.}
\label{fig:sersic}
\end{center}
\end{figure}

As shown in Figure~\ref{fig:sersic}, the presence of an AGN contributing less than 30 per cent of the rest-frame $B$ band light affects $n$ approximately as expected for most of the galaxies considered. The median increase is weaker among disc-dominated systems than among bulge-dominated or bulge$+$disc systems, though the maximum changes are greatest among the disc-dominated systems, suggesting that the fraction of AGN host galaxies that are classified as disc-dominated should be considered a lower limit to the true fraction of disc-dominated AGN hosts. Similarly, the fraction of AGN host galaxies classified as bulge-dominated should be considered an upper limit.

Simmons \& Urry (2008) measured the change $\Delta n$ for a large sample of real and simulated galaxies with point sources added. For input point source magnitudes $M > 22$, they found that GALFIT is generally able to recover the correct $n$, with an apparently smaller scatter than what is presented here. Differences between our results and those of Simmons \& Urry (2008) may be partially explained by the differences in the sample sizes. While the initial sample presented here contains 622 galaxies, that presented by Simmons \& Urry (2008) contains 12,592 galaxies, providing more statistically significant results. In addition, they combined results from all galaxy types, which may hide the differences shown in Figure~\ref{fig:sersic}.

\subsubsection{Combined morphology measurements}\label{expt_1:combined_morphs}
\begin{figure}
\begin{center}
\includegraphics[width=3.5in]{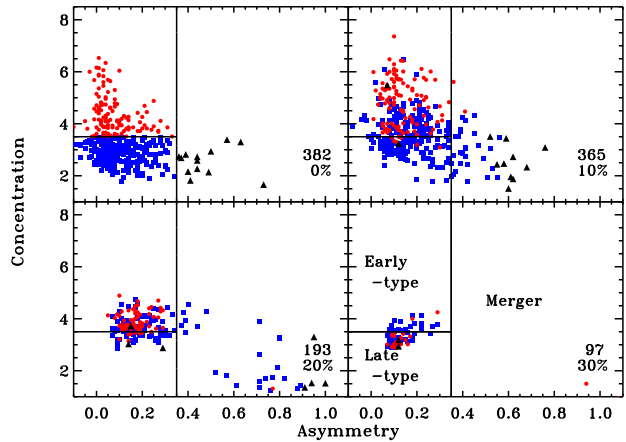}
\caption{$C$-$A$ diagrams of simulated AGN host galaxies. The values in the lower right corners of each panel indicate the fraction of rest-frame $B$ band light contributed by the AGNs represented in the panel and the number of host galaxies for which we have reliable morphologies; only those systems that meet the reliability criteria are included. Solid lines separate the galaxies by morphology classification, as indicated on the final panel. The symbol shapes and colours indicate the initial $C$-$A$ morphologies, following Figure~\ref{fig:asym}.}
\label{fig:conc_asym}
\end{center}
\end{figure}

Figure~\ref{fig:conc_asym} shows the results of morphology classifications based on a combination of $C$, which is expected to increase as the AGN fraction increases toward $\%_{\rm AGN} =$ 20 per cent, and $A$, which is expected to experience minimal change, unless the AGN is offset from the galaxy centre. At $\%_{\rm AGN} =$ 10 per cent, more than 90 per cent (106/116) of the galaxies initially classified as early-type retain their classification. In contrast, only 41 per cent (96/237) of the late-type galaxies are still classified as such, and almost half (46 per cent, 110/237) are classified as early-type. Galaxies initially classified as mergers tend to retain that classification at $\%_{\rm AGN} =$ 10 per cent; the seven mergers with reliable morphologies at $\%_{\rm AGN} =$ 20 per cent are split between mergers (four), early-type (one), and late-type (two).

The $C$-$A$ classification method has the advantage that few of the early-type galaxies are incorrectly classified as the AGN fraction increases. However, the early-type classification becomes significantly contaminated with late-type galaxies, and $C$ measurements are unreliable for AGN fractions $\%_{\rm AGN} >$ 20 per cent. For AGN hosts, the early-type classification is rather complete, but impure, while the late-type classification is mostly pure, but incomplete; the merger classification is contaminated with late-type galaxies and incomplete due to a decrease in the asymmetry of a few galaxies.

\begin{figure}
\begin{center}
\includegraphics[width=3.5in]{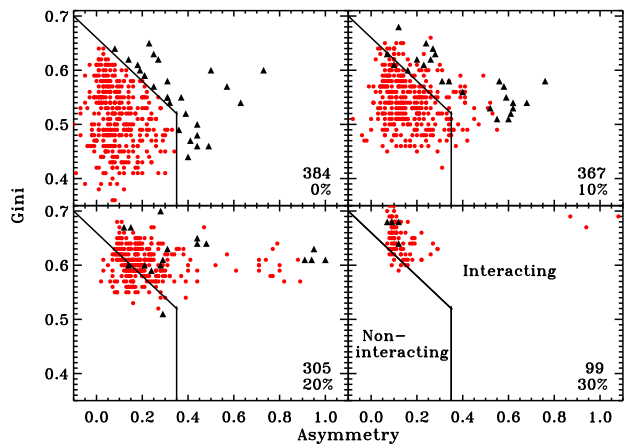}
\caption{$G$-$A$ diagrams of simulated AGN host galaxies. The layout of this figure follows Figure~\ref{fig:conc_asym}. Galaxies initially classified as interacting (non-interacting) by $G$-$A$ are represented by black triangles (red circles).}
\label{fig:gini_asym}
\end{center}
\end{figure}

In Figure~\ref{fig:gini_asym} we demonstrate the results of pairing $G$ and $A$ to classify galaxies as interacting or non-interacting. As previously discussed, $A$ changes little with increasing $\%_{\rm AGN}$ unless the AGN is located away from the geometric centre of the galaxy, so the main result shown in this figure is due to changes in $G$, which tends to increase with $\%_{\rm AGN}$ (cf.\ Figure~\ref{fig:gini}).

Among galaxies initially classified as interacting by the $G$-$A$ method (black triangles in Figure~\ref{fig:gini_asym}), very few are misclassified as non-interacting as the fraction of light coming from the point source increases; as a result, the non-interacting classification remains pure. However, as the AGN contribution increases, the region indicating interaction suffers from contamination by misclassified non-interacting galaxies. For $\%_{\rm AGN} =$ 10 per cent, 21 per cent (72/343) of the non-interacting galaxies are instead classified as interacting; for $\%_{\rm AGN} =$ 20 per cent, this fraction increases to 71 per cent (204/288). Thus, the fraction of AGN hosts classified by $G$-$A$ as interacting represents an upper limit, and the fraction of AGN host galaxies classified as non-interacting should be considered a lower limit.

\begin{figure}
\begin{center}
\includegraphics[width=3.5in]{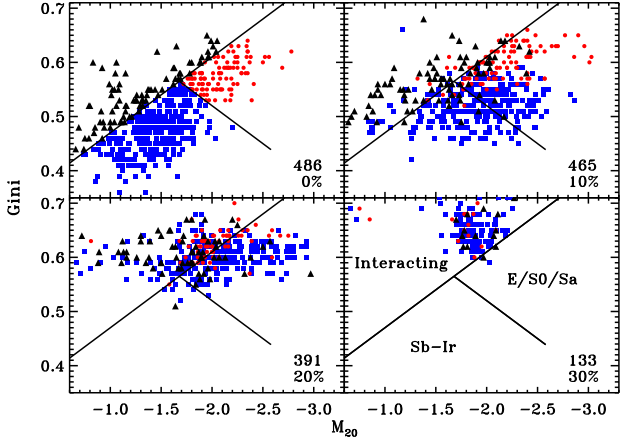}
\caption{$G$-$M_{20}$ diagrams of simulated AGN host galaxies. The layout of this figure follows Figure~\ref{fig:conc_asym}. Symbols indicate the initial morphologies, following Figure~\ref{fig:gini}.}
\label{fig:gini_m20}
\end{center}
\end{figure}

As a galaxy's nuclear flux increases, the measurement of the distribution of light, $G$, is expected to monotonically increase because more of the light is distributed among fewer of the pixels associated with the galaxy. In contrast, the value of $M_{20}$ is expected to {\it decrease} as the fraction of light contributed by an active nucleus increases. Increasing the brightness of a few adjacent pixels, such as within the nuclear region of a galaxy, tends to decrease the spatial distribution of the brightest 20 per cent of the pixels without changing the overall spatial distribution of the pixels, thereby decreasing $M_{20}$ relative to its initial value. Together, these effects are expected to shift $G$-$M_{20}$ measurements toward the E/S0/Sa classification (\S~\ref{expt_1:morph_class}). Results from the pairing of $G$ with $M_{20}$ are presented in Figure~\ref{fig:gini_m20}.

Sb-Ir galaxies are significantly affected by the presence of an AGN. At $\%_{\rm AGN} =$ 10 per cent, only 46 per cent (132/288) of the initially Sb-Ir galaxies that still meet the reliability criteria are again classified as Sb-Ir, 46 per cent are classified as E/S0/Sa, and 8 per cent are classified as interacting. In contrast, 81 per cent (78/96) of the galaxies initially classified as E/S0/Sa, and considered to have reliable morphologies, are `correctly' classified at $\%_{\rm AGN} =$ 10 per cent, while 4 per cent (4/96) are classified as Sb-Ir and 14 per cent (13/96) are classified as interacting. However, at $\%_{\rm AGN} =$ 20 per cent, only 34 per cent (23/68) of the galaxies initially classified as E/S0/Sa are still so classified; due to increases in $G$, most are instead classified as interacting. The galaxies initially classified as interacting tend to retain that classification.

\begin{figure}
\begin{center}
\includegraphics[width=3.5in]{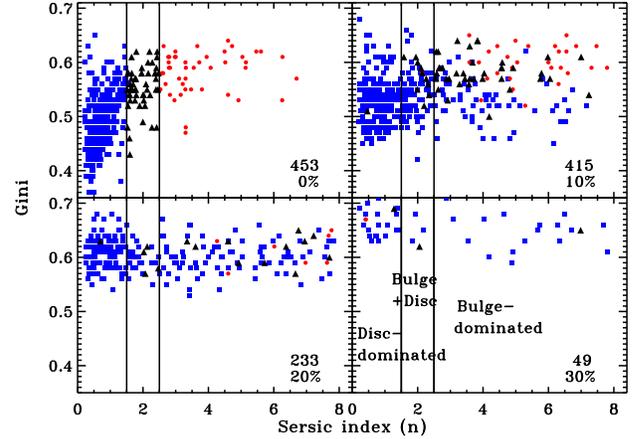}
\caption{$G$-\sersic\/ index diagrams of simulated AGN host galaxies. The layout of this figure follows Figure~\ref{fig:conc_asym}. The symbol shapes and colours indicate the initial \sersic\/ profiles, following Figure~\ref{fig:sersic}.}
\label{fig:gini_sersic}
\end{center}
\end{figure}

\begin{figure}
\begin{center}
\includegraphics[width=3.5in]{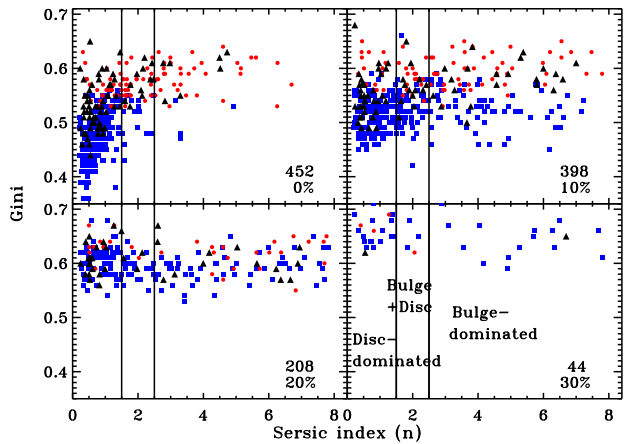}
\caption{$G$-\sersic\/ index diagrams of simulated AGN host galaxies. The layout of this figure follows Figure~\ref{fig:conc_asym}. The symbol shapes and colours indicate the initial $G$-$M_{20}$ morphologies, following Figure~\ref{fig:gini}.}
\label{fig:gini_sersic_gm20}
\end{center}
\end{figure}

Both $G$ and the \sersic\/ index $n$ are expected to increase with an increase in the AGN fraction; Figures~\ref{fig:gini_sersic} and~\ref{fig:gini_sersic_gm20} depict the pairing of these two measurements. Symbols in the former indicate the original \sersic\/ profiles, while symbols in the latter indicate the original $G$-$M_{20}$ morphology classifications. The number of objects represented in the two figures differs because the set of galaxies that initially meet the \sersic\/ index reliability criteria does not completely overlap with the set of galaxies that initially meet the $G$-$M_{20}$ reliability criteria.

The addition of a point source tends to increase the value of $G$ so that although the galaxy sample presented in Figure~\ref{fig:gini_sersic} initially has a large range of $G$ values, the range for $\%_{\rm AGN} \ge$ 20 per cent is rather narrow. The \sersic\/ index also tends to increase as the fraction of light coming from an added point source is increased, but this trend does not apply equally to all galaxies, as indicated by the wide spread in \sersic\/ indices at high AGN fractions. This suggests that $G$ is more strongly affected than the \sersic\/ index by the presence of a point source in a galaxy.

At $\%_{\rm AGN} =$ 10 per cent, 58 per cent (230/395) of the initially disc-dominated galaxies retain that classification, and at $\%_{\rm AGN} =$ 20 per cent, 51 per cent (133/260) are still correctly classified. AGNs also significantly affect the \sersic\/ profiles of systems initially classified as Sb-Ir galaxies, as demonstrated by Figure~\ref{fig:gini_sersic_gm20}. As the AGN fraction increases, about half of the systems initially classified as Sb-Ir galaxies transition from disc-dominated or bulge$+$disc classifications into bulge-dominated classifications. The bulge$+$disc galaxy \sersic\/ indices are more easily increased, with only 21 per cent (12/56) still classified as bulge$+$disc at $\%_{\rm AGN} =$ 10 per cent. As a result of these two trends, the bulge-dominated classification is quickly contaminated. The fraction of AGN host galaxies classified as bulge-dominated should be considered an upper limit, while the reverse is true for the fractions of AGN hosts classified as disc-dominated or bulge$+$disc.

\subsection{Summary of morphology results}\label{expt_1:summary}
With this first experiment, we have investigated the effect of an AGN on the morphological measurements $G$, $M_{20}$, $C$, $A$, and the \sersic\/ index. First, we find that the number of galaxies deemed to have reliable morphology measurements decreases significantly for $\%_{\rm AGN} \ge$ 20 per cent, due almost exclusively to an elliptical Petrosian radius smaller than r$_{\rm ell} = 0.3\arcsec$.

$C$ and $M_{20}$ are adversely affected for AGN fractions $\%_{\rm AGN} \ge$ 20 per cent on account of their dependence on the central and/or brightest 20 per cent of a galaxy's light. In such cases, the morphology measurements are effectively measuring the light from the AGN, with minimal input from the rest of the galaxy. Thus, $C$ and $M_{20}$ are expected to be unreliable when classifying the morphologies of galaxies hosting bright, unobscured AGNs.

The Gini coefficient of Sb-Ir galaxies is significantly affected even for an AGN fraction of 10 per cent, with an estimated 54 per cent misclassified as E/S0/Sa or interacting galaxies. For the same AGN fraction, the \sersic\/ index is slightly less affected by the presence of an AGN; 42 per cent of the disc-dominated host galaxies are misclassified as bulge$+$disc or bulge-dominated.

However, not all of the morphology measurements are so adversely affected. $A$ experiences little change as the AGN fraction increases, unless the location of the AGN is offset from the geometrical centre of the galaxy. Thus, $A$ may still reliably distinguish between interacting/merging and non-interacting/non-merging galaxies.

The end result is that an AGN contributing more than 10-20 per cent of a system's light in a given band (here, the rest-frame $B$ band) significantly affects most morphology measurements. However, disc galaxies hosting AGNs contributing more than 20 per cent of the $B$ band light may be visually identifiable (\S~\ref{expt_1:visual_agn}), alerting researchers to take special care with such systems. At the highest AGN fractions, AGNs hosted by elliptical galaxies appear to have sharper edges than typical stellar bulges, thus they may be visually identifiable in certain systems. Visual inspection of a sub-sample of the simulated galaxies described here suggests that about half of the AGNs are visually identifiable for AGN fractions greater than 20-30 per cent, but for a few unusual cases, the AGNs are not clearly distinguishable from a nuclear starburst even at the highest AGN fractions.

%%%%%%%%%%%%%%%%%%%%%%%%%%%%%
\section{Experiment \#2: Combining AGN and Non-AGN Galaxy Spectral Templates}\label{expt_2}

For the second experiment, an investigation of the effect of an AGN on measured optical and UV-optical colours, a spectral template representing a quasar is added to spectral templates representing three non-AGN galaxies (an elliptical, an Sb galaxy, and a starburst galaxy). The quasar template is scaled to contribute specified fractions of the rest-frame $B$ band flux of the resulting galaxy/AGN systems, consistent with the method used for the first experiment (\S~\ref{expt_1}). We then measure the rest-frame optical and UV-optical colours of the resulting systems. The goal of this experiment is to determine the fraction of the total flux contributed by an AGN that is necessary to move a red sequence or green valley galaxy into the green valley or the blue cloud, respectively.

\subsection{Spectroscopic simulation of an AGN}\label{expt_2:spec_agn}
Before combining the spectral templates described in \S\S~\ref{data:agn_template} and~\ref{data:quiescent_template}, the rest-frame $B$ band flux from the quasar template is scaled to contribute specified fractions (5 per cent, 10 per cent, 15 per cent, ... 50 per cent) of the rest-frame $B$ band flux from the resulting galaxy$+$quasar spectra. After adding the scaled quasar spectrum to the galaxy spectra, the NUV$-r$ and $U-B$ colours of each of the original and new systems is measured, and the fraction of light from the AGN spectra that moved a red sequence or green valley galaxy into the green valley or blue cloud is determined.

Synphot\footnote{http://www.stsci.edu/resources/software\_hardware/stsdas\\/synphot}, a package available for the Image Reduction and Analysis Facility\footnote{http://iraf.noao.edu/iraf-homepage.html} includes a task named \verb|calcphot| that measures the flux $f_{\nu}$ [erg s$^{-1}$ cm$^{-2}$ Hz$^{-1}$] in a specified passband for an input spectrum. Using this task we measure $f_{\nu}$ for the GALEX NUV, SDSS $r$, and Johnson $U$ and $B$ passbands from each of the spectral templates. Scaling of the AGN templates with respect to the non-AGN galaxy templates is based on the $B$ passband fluxes measured from each spectral template as follows:
\begin{equation}
f_{X_{\rm tot}} = f_{X_{\rm gal}} + \frac{f_{X_{\rm AGN}} \times f_{B_{\rm gal}}}{f_{B_{\rm AGN}}} \times \frac{\%_{\rm AGN}}{1 - \%_{\rm AGN}};
\end{equation}
where $\%_{\rm AGN}$ represents the specified fraction of the total $B$ band flux contributed by the AGN spectral template; $f_{B_{\rm tot/gal/AGN}}$ represents the $B$ band flux contributed by the combined system, the galaxy, or the AGN; and $f_{X_{\rm tot/gal/AGN}}$ represents the NUV, $r$, or $U$ flux contributed by the combined system, the galaxy, or the AGN. The NUV$-r$ and $U-B$ colours are then calculated as follows:
\begin{equation}
{\rm NUV} - r = -2.5 \times \log(f_{\rm NUV}/f_{r});
\end{equation}
\begin{equation}
U - B = -2.5 \times \log(f_{U}/f_{B}).
\end{equation}
This method of calculating the colour is equivalent to first converting the fluxes to AB magnitudes [e.g., NUV $= −2.5 \times \log(f_{\rm NUV}) + 48.60$] and then calculating the difference between the NUV and $r$ or $U$ and $B$ measurements.

\begin{figure}
\begin{center}
\includegraphics[width=3.5in]{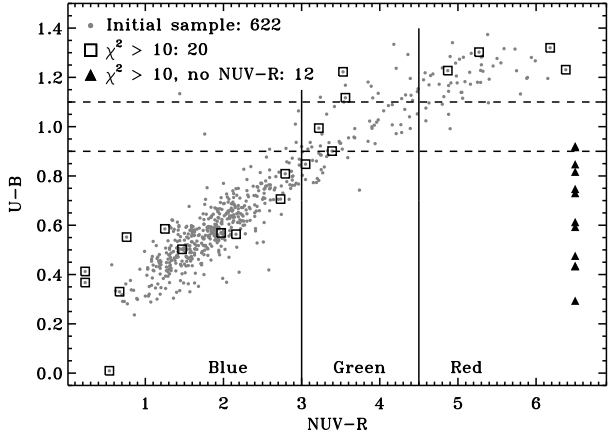}
\caption{$U-B$ vs.\ NUV$-R$ of the initial galaxy sample described in \S~\ref{data:images}. [The colours represented in this figure come from different data (\S~\ref{data:images}) than the colours presented in the next few figures (\S\S~\ref{data:agn_template} and~\ref{data:quiescent_template}), so the UV-optical colours shown here (NUV$-R$) differ slightly from the UV-optical colours shown in the later figures (NUV$-r$).] Gray dots represent the available rest-frame optical and UV-optical colours. Galaxies for which UV-optical colours are available but not considered reliable are marked with a black square, and those for which UV-optical colours are unavailable are represented by black triangles and set to have NUV$-R = 6.5$. Solid lines indicate the divisions between blue, green, and red UV-optical colours, as labeled; dashed lines indicate approximate divisions between blue ($U-B < 0.9$) and red ($U-B > 1.1$) optical colours. Compared to the UV-optical colours, the optical colours saturate at $U-B \sim 1.2$.}
\label{fig:ub_nuvr}
\end{center}
\end{figure}

Roughly following Willmer et al.\ (2006), Nandra et al.\ (2007), and Weiner et al.\ (2009), we define red and blue optical colours as $U-B > 1.1$ and $U-B < 0.9$, respectively. The red and blue regions defined in the references are based on a combination of $U-B$ and $M_{B}$, but since we do not use $M_{B}$ for the current study, we approximate the divisions between regions using only the $U-B$ colour. Blue, green, and red UV-optical colours are defined as NUV$-r < 3$, $3 < $NUV$-r < 4.5$, and NUV$-r > 4.5$, respectively, following, e.g., Wyder et al.\ (2007). The UV-optical green valley encompasses the significant population exhibiting colours between the UV-optical red sequence and blue cloud; a similarly significant population is {\it not} observed between the {\it optical} red and blue regions. Figure~\ref{fig:ub_nuvr} demonstrates the general consistency between the optical and UV-optical colours of the initial sample described in \S~\ref{data:images}, as well as the divisions between colours. The lack of a significant {\it optical} green valley population is evident from this figure.

\subsection{Results}\label{expt_2:results}
\begin{figure}
\begin{center}
\includegraphics[width=3.5in]{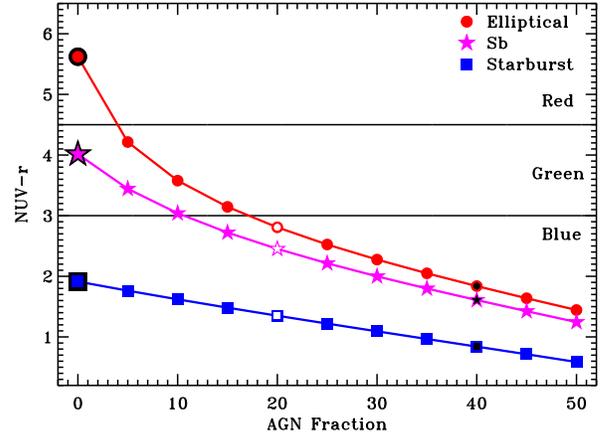}
\caption{NUV$-r$ as a function of the fraction of $B$ band flux contributed by an AGN. Red circles represent the elliptical template, magenta stars represent the Sb template, and blue squares represent the starburst template. Symbols outlined in black represent galaxies at $\%_{\rm AGN} = 0$ per cent, symbols filled with white represent galaxies at $\%_{\rm AGN} =$ 20 per cent, and symbols filled with black represent galaxies at $\%_{\rm AGN} =$ 40 per cent. Horizontal lines indicate the separations between red, green, and blue UV-optical colours. The AGN template has a colour NUV$-r = -1.09$.}
\label{fig:nuvr_psf}
\end{center}
\end{figure}

Figure~\ref{fig:nuvr_psf} shows the change in NUV$-r$ colour as a function of the fraction of $B$ band flux contributed by the quasar template. The colours of the systems resulting from the combination of the AGN and non-AGN galaxy templates tend toward the colour of the AGN, NUV$-r = -1.09$, as the AGN fraction increases.

The quasar template brings the initially red elliptical galaxy down into the green valley and then the blue cloud by AGN fractions of 5 per cent and 20 per cent, respectively. The Sb galaxy is also drawn from its initial location in the green valley down to the blue cloud when only 10 per cent of the $B$ band flux is contributed by the AGN. Although the colour of the starburst galaxy is also affected by the quasar template, it is blue to begin with, so the effect is less significant.

These results suggest that the UV-optical colours of AGN host galaxies may be more significantly affected by an active nucleus than previous work has indicated (e.g., Kauffmann et al.\ 2007). However, Kauffmann et al.\ (2007) specifically excluded galaxies exhibiting quasar-like spectra (e.g., AGNs typically associated with a central point source), so this might be a result that depends very strongly on the type of AGN observed. In particular, AGNs that do not exhibit a rising blue continuum, such as LINERs and Type-2 Seyfert galaxies, are not expected to affect the measured host galaxy colours as significantly as the results shown here.

\begin{figure}
\begin{center}
\includegraphics[width=3.5in]{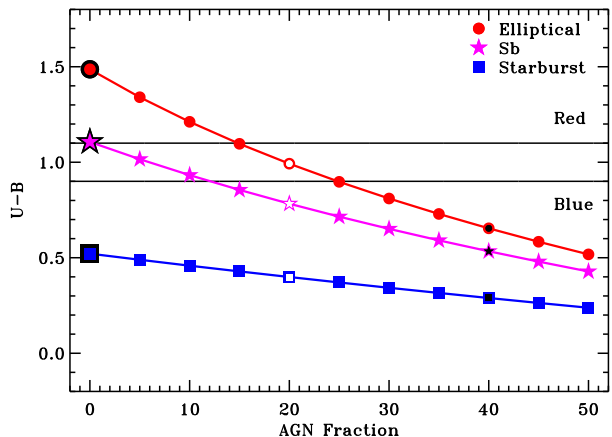}
\caption{$U-B$ as a function of the fraction of $B$ band flux contributed by an AGN. Symbols are as for Figure~\ref{fig:nuvr_psf}. Horizontal lines indicate approximate separations between red and blue optical colours. The AGN template has a colour $U-B = -0.01$.}
\label{fig:ub_psf}
\end{center}
\end{figure}

The AGN template similarly affects the $U-B$ colours of the three non-AGN galaxies, as Figure~\ref{fig:ub_psf} demonstrates. Again, we find that the quasar template significantly affects measurements of both the elliptical and the Sb galaxy template colours, bringing them down into the blue cloud by AGN fractions of 25 per cent 15 per cent, respectively.

Comparing Figure~\ref{fig:ub_psf} to results from Nandra et al.\ (2007; see their Figure 1), we conclude that observed AGN host galaxies that exhibit colours $U-B < 0.5$ may be consistent with optically red galaxies, the light from which has been contaminated by AGNs contributing more than 50 per cent of the system's rest-frame $B$ band light. Their estimate of AGN-caused colour contamination (solid line on their Figure 1) is consistent with our result for an AGN fraction of 25-30 per cent.

It is clear that the light from AGN host galaxies exhibiting colours $U-B < 0.5$ (e.g., Nandra et al.\ 2007), is most likely contaminated by the quasar light. However, it is less obvious what to conclude about the AGN hosts located in the upper region of the blue cloud. Pierce et al.\ (2010) address this further by considering the relationships between host galaxy aperture colours and the AGN X-ray luminosities and hardness ratios.

\begin{figure}
\begin{center}
\includegraphics[width=3.5in]{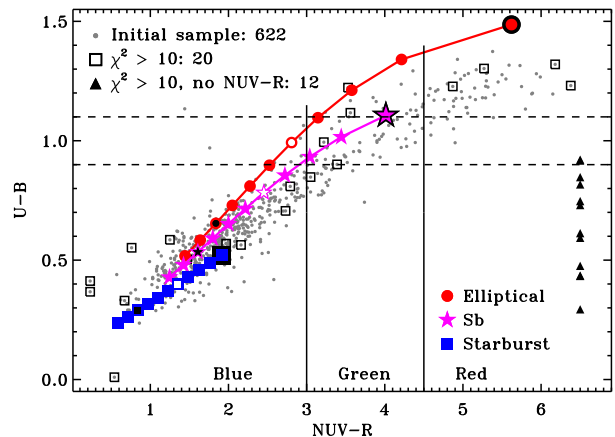}
\caption{$U-B$ vs.\ NUV$-R$ of the galaxy samples and templates described in \S~\ref{data}. Small (black and gray) symbols follow Figure~\ref{fig:ub_nuvr}. Large (coloured) symbols follow Figure~\ref{fig:nuvr_psf}. For most AGN fractions, the observed colours appear consistent with colours of galaxies not hosting AGNs.}
\label{fig:ub_nuvr_psf}
\end{center}
\end{figure}

Finally, Figure~\ref{fig:ub_nuvr_psf} again shows the optical and UV-optical colours of the galaxy sample shown in Figure~\ref{fig:ub_nuvr} (small gray and black symbols), but it also includes the AGN colour tracks created by combining Figures~\ref{fig:nuvr_psf} and~\ref{fig:ub_psf}. Regardless of AGN fraction, the colours of the Sb$+$AGN template (magenta stars) are similar to the rest-frame colours of observed galaxies, while the starburst$+$AGN template colours (blue squares) are consistent with the colours of observed galaxies for all but the highest AGN fractions. In contrast, the elliptical$+$AGN template (red circles) is optically redder than most of the observed galaxies suggesting that such a colour-colour diagram of AGN host galaxies (particularly elliptical galaxies) may facilitate the identification of colour contamination. Contaminated AGN host galaxies in the UV-optical green valley may exhibit unusually red {\it optical} colours.

%%%%%%%%%%%%%%%%%%%%%%%%%%%%%
\section{Colours vs.\ S\'{e}rsic Indices of Simulated AGN Host Galaxies}\label{nuvr_morph}
Observations indicate a general correlation between the colour and the morphology of an undisturbed galaxy (e.g., B$\ddot{o}$hm \& Wisotzki 2007; Lotz et al.\ 2008a; Silverman et al.\ 2008), which suggests that unusual colour-morphology combinations among AGN host galaxies (e.g., blue ellipticals) may indicate examples of AGN contamination. In addition, Lotz et al.\ (2008b) showed that it is possible to use the recent star formation history and morphology of a {\it merging} galaxy to determine the observed merger stage. This information, combined with observations such as X-ray luminosity and obscuration, facilitates the study of AGN lifetimes and the possible connections between galaxy interactions and the initiation of supermassive black hole growth.

In this section, we present colour-morphology diagrams created by matching results from the optical and UV-optical colour simulations (\S~\ref{expt_2}) to results from the morphology simulations (\S~\ref{expt_1:results}). For each AGN fraction, we match colour results from the quasar$+$elliptical, quasar$+$Sb, and quasar$+$starburst templates to the median morphologies of galaxies initially classified as having bulge-dominated, bulge$+$disc, and disc-dominated \sersic\/ profiles, respectively.

\begin{figure}
\begin{center}
\includegraphics[width=3.5in]{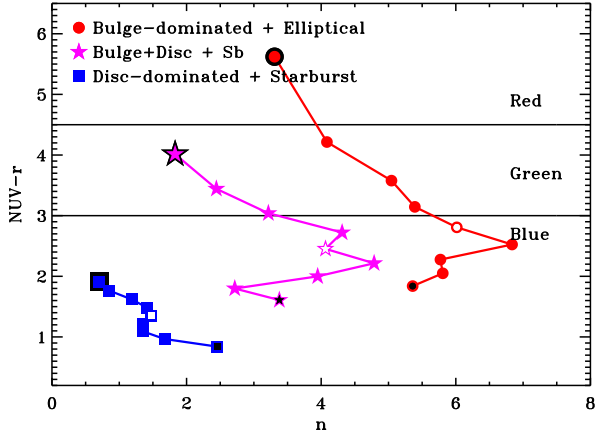}
\caption{NUV$-r$ vs.\ \sersic\/ index of non-AGN galaxies with AGNs added. Symbols and lines follow Figure~\ref{fig:nuvr_psf}, except that the symbols also represent galaxies initially classified as bulge-dominated (red circles), bulge$+$disc (magenta stars), or disc-dominated (blue squares). Horizontal lines divide the diagram into red, green, and blue UV-optical colours, as defined in \S~\ref{expt_2}.}
\label{fig:nuvr_sersic}
\end{center}
\end{figure}

Figure~\ref{fig:nuvr_sersic}, a plot of \sersic\/ index vs.\ UV-optical colour, indicates a general anti-correlation between colour and \sersic\/ index. As the fraction of light contributed by an AGN increases to 25 per cent, the \sersic\/ index generally {\it increases}, while the colour becomes bluer. For AGN fractions exceeding 25-30 per cent, the \sersic\/ indices of bulge-dominated and bulge$+$disc systems are found to {\it decrease} as the colour becomes bluer, suggesting a reliability limit for systems with strong point sources. This figure also suggests that AGN hosts with \sersic\/ indices $n > 2.5$ and NUV$-r < 2.5$ are strong candidates for high levels of AGN contamination (e.g., $\%_{\rm AGN} >$ 20 per cent).

Few X-ray-selected AGNs are observed with colour-morphology characteristics similar to those of the strongly contaminated systems (e.g., Pierce 2009), suggesting either that most X-ray-selected AGNs contribute $<$ 20 per cent of the rest-frame $B$ band light detected from the system or that the contributed AGN light is significantly reddened, in contrast to the quasar template used here.

At AGN fractions less than 20 per cent, there is reasonable concern that some AGN host galaxies are incorrectly classified as green valley galaxies {\it and not easily identified as being problematic} because of the low AGN fraction; thus extra caution should be taken when studying such systems. Using aperture photometry to separate the inner and outer colours of the AGN hosts can help with this, as discussed by Ammons et al.\ (in preparation) and Pierce et al.\ (2010).

\begin{figure}
\begin{center}
\includegraphics[width=3.5in]{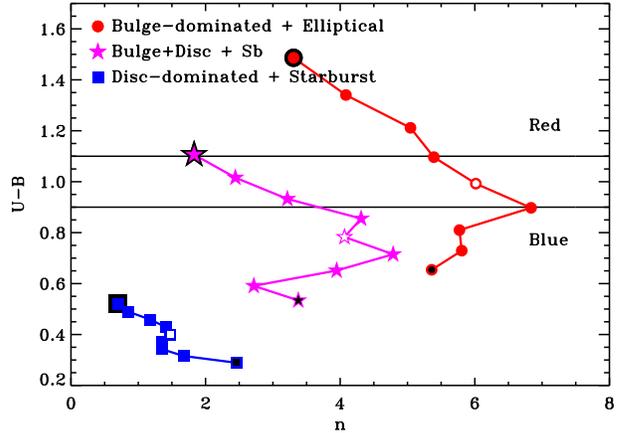}
\caption{$U-B$ vs.\ \sersic\/ index of non-AGN galaxies with AGNs added. Lines and symbols follow Figure~\ref{fig:nuvr_sersic}, except that horizontal lines divide the diagram into red and blue optical colours (e.g., Willmer et al.\ 2006).}
\label{fig:ub_sersic}
\end{center}
\end{figure}

We also compare the \sersic\/ indices to the $U-B$ colours of our simulated systems, as presented in Figure~\ref{fig:ub_sersic}. $U-B$ {\it appears} less sensitive than NUV$-r$ to the contribution from an unobscured AGN, in particular for the initially bulge-dominated system (red circles). This system falls in the middle of the optical `green valley' at an AGN fraction of 20 per cent, in contrast to its UV-optical blue cloud classification at the same AGN fraction. However, with respect to the green valley, optical colours are less discriminating than UV-optical colours, as noted in \S~\ref{expt_2:results}.

%%%%%%%%%%%%%%%%%%%%%%%%%%%%%
\section{Discussion and Summary}\label{summary}
It is reasonable to expect that galaxies hosting unobscured AGNs (i.e., those visible as point sources) may experience significant contamination of their morphological and colour measurements. This paper describes our method for and results from testing the reliability of common measures of galaxy morphology and colour when an AGN contributes a specified fraction of the observed $B$ band light. We emphasize that the work presented here was specifically designed to investigate the limitations of various morphology and colour measurements in the presence of an AGN, in part for the sake of better understanding previous results that relied on these measurements without explicitly accounting for the AGN light. Others have carefully investigated AGN host galaxy morphologies and colours, explicitly removing the AGN contribution (e.g., S\'{a}nchez et al.\ 2008; Gabor et al.\ 2009), and the reliability of host galaxy/AGN decompositions with respect to PSF variations (Kim et al.\ 2004).

For the first experiment (\S\S~\ref{expt_1},~\ref{expt_1:results}), we simulate AGNs, represented by a $V$ band PSF, in a sample of AEGIS galaxies not known to host AGNs. We first find that increasing the fraction of observed light contributed by an AGN decreases the measured elliptical Petrosian radius, which in turn decreases the number of galaxies for which we can determine reliable morphologies (\S~\ref{expt_1}). This decrease is significant for AGN fractions exceeding 20 per cent. Two of the morphology measurements, $C$ and $M_{20}$, are almost certainly unreliable for AGNs that contribute more than 20 per cent of the galaxy's $B$ band light. By definition, these two measurements depend in part on locating either the central or the brightest 20 per cent of the flux; if that flux comes directly from an AGN, the measurements focus on the AGN and may not properly characterize the host galaxy. The presence of an AGN also detrimentally affects measurements of the \sersic\/ index and $G$, as expected, though the effects are not inherently unreliable. In contrast, unless the AGN is offset from the geometric centre of the galaxy, $A$ remains generally reliable even for high AGN fractions. As described in \S~\ref{expt_1:results}, the results from a similar experiment by Simmons \& Urry (2008) are complementary to the results presented here.

Gabor et al.\ (2009) showed that AGN host galaxy concentrations are lower when the AGN contribution is first fitted with a PSF and then {\it subtracted} from the image, consistent with our results that the {\it addition} of a point source tends to increase the measured concentration. However, their result that the asymmetry of a galaxy may be artificially low if the AGN is not properly accounted for does not seem consistent with the results presented here. Residuals shown by Kim et al.\ (2008) suggest that PSF subtraction may sometimes increase the apparent asymmetry of the galaxy, unless the PSF is very well matched to the AGN being fitted. Though the current work uses only a single PSF for all of the simulations, we do not attempt to subtract it from the images; in addition, we provide GALFIT with the same PSF that we add to the galaxy images. Thus we do not expect to run into problems with this particular issue.

For the second experiment (\S~\ref{expt_2}), we add a quasar spectral template to three non-AGN spectral templates to determine how an AGN affects the optical and UV-optical colours of its host galaxy. The quasar spectral template is scaled to contribute a set of specified fractions of the rest-frame $B$ band light emanating from the final galaxy/AGN systems, and the colours of the original and final systems are measured and compared. The use of a quasar spectral template avoids the complications which would be introduced by the use of a spectral template representing an obscured AGN, such as the effect of the obscuring gas on the AGN contribution to the colour of the system. However, the quasar template represents an extreme example of an AGN, meaning that our results regarding the potential colour biases are to be considered upper limits.

Quasar light strongly affects the measured UV-optical colours of the simulated quasar host galaxies. A quasar contributing as little as 5 per cent of the observed $B$ band light can move the observed NUV$-r$ colour of an elliptical galaxy from the UV-optical red sequence to the green valley, and an AGN fraction of 20 per cent moves it into the blue cloud. Optical colours are similarly affected. An AGN fraction of 25 per cent shifts the observed optical colour of the elliptical galaxy from the red sequence into the blue cloud. An encouraging caveat is that there are relatively few AGNs that have such unobscured quasar spectra, and we expect that such systems would be easily identified by visual inspection of the galaxy images.

S\'{a}nchez et al.\ (2004), who studied a sample of 15 optically selected, intermediate luminosity, type 1 AGNs, found that the colours of the early-type galaxies, which comprised much of their sample, tended to be on the blue edge of the red sequence, if not in the blue cloud itself. Having first fitted the AGNs with PSFs and subtracted them from the images, these colours are expected to accurately represent the galaxies without contamination from the AGN light. In contrast, the artificially blue colours that result from the combination of a QSO spectrum and a quiescent galaxy spectrum, as shown here, represent an upper limit to the contamination from AGN light.

By combining the results from our two experiments (\S~\ref{nuvr_morph}), we find that systems for which an AGN contributes more than 20 per cent of the $B$ band light appear significantly bluer and more bulge-dominated than the underlying galaxy. Therefore, we would expect that AGNs hosted by apparently blue, bulge-dominated galaxies bias the colour and morphology measurements. Pierce (2009) found that such systems make up less than half of the AGN population at $z \sim 1$, and many of the AGNs that do fit that description have visible point sources. Red, bulge-dominated AGN host galaxies are not expected to experience contamination from the AGN light.

Based on an inspection of our simulated galaxies, it appears that the point sources in such systems should be visually identifiable for high AGN fractions. The results of Pierce et al.\ (2010), in which the authors compare the observed nuclear, extended, and integrated optical colours of galaxies hosting X-ray-selected AGNs to a corresponding control sample, suggests a low incidence of systems with AGN fractions $\%_{\rm AGN} \ge$ 20 per cent because the integrated colours of most of the AGN host galaxies correlate much more closely with the extended colours than with the nuclear colours.

Employing a collection of simulations, the current study investigates the extent to which various methods used to measure galaxy colours and morphologies may be biased by the presence of an AGN. This is particularly relevant when (1) considering the significance and timing of the influence that an AGN may exert on the star formation rate of its host galaxy (i.e., are `green valley' AGN host galaxies intrinsically red?) and (2) using the morphological characteristics of AGN host galaxies to investigate relationships between interactions and the activation of the central supermassive black hole (i.e., could the many AGN host galaxies classified as bulge-dominated systems actually have disc-like or disturbed morphologies which are hidden because of light from the nuclear regions?).

Our results suggest that the answer to the second question is tentatively `no', unless there is a visually identifiable point source. The answer to the first question is less certain, because we find that the colours may be affected by even a small AGN fraction. However, the quasar template used here represents an extreme (and relatively rare) case of an unobscured AGN; an increase in the obscuration of the AGN is expected to decrease the colour contamination (cf.\ Figure 1 of Nandra et al.\ 2007). Therefore, our results are not necessarily inconsistent with earlier results (e.g., Kauffmann et al.\ 2007), which indicate that galaxies hosting the more numerous AGNs (e.g., Seyfert galaxies and LINERs) are not likely to suffer significant contamination from the nuclear regions.

Measurements of AGN host galaxies that indicate bulge-dominated \sersic\/ profiles and blue colours are not necessarily biased by light from the AGN. Studying the aperture colours of AGN host galaxies should help to resolve the question of the origin of the blue colour. Pierce et al.\ (2010) find that X-ray-selected AGNs hosted by UV-optical green valley galaxies typically exhibit extended optical colours that are similar to the extended colours of a control sample, which suggests that most green valley galaxies hosting AGNs are intrinsically green and that it is rare for red sequence galaxies to host AGNs contributing more than $\sim 10$ per cent of the $B$ band light.

\section*{Acknowledgments}
C.\ M.\ P.\ and J.\ R.\ P.\ acknowledge support from NASA ATP grant NNX07AG94G. C.\ M.\ P.\ additionally acknowledges support from the Departments of Physics at the University of California, Santa Cruz and the Georgia Institute of Technology, and from the DEEP2 survey. J.\ M.\ L.\ acknowledges support from the NOAO Leo Goldberg Fellowship and NASA grant HST-AR-9998. We also thank the referee for comments that helped improve this manuscript. This study makes use of data from AEGIS, a multi-wavelength sky survey conducted with the {\it Chandra}, GALEX, {\it Hubble}, Keck, CFHT, MMT, Subaru, Palomar, {\it Spitzer}, VLA, and other telescopes and supported in part by the NSF, NASA, and the STFC. The Advanced Camera for Surveys General Catalog (ACS-GC) data set (Griffith et al., in preparation) has also been used for this study. In addition, this research has made use of the NASA/IPAC Extragalactic Database (NED) which is operated by the Jet Propulsion Laboratory, California Institute of Technology, under contract with the National Aeronautics and Space Administration.

\bsp

\label{lastpage}

\end{document}